\documentstyle[12pt,aaspp4,eps]{article}
\def\kms{km\thinspace s$^{-1}$}                  
\begin{document}
\title{SBF Distances to Dwarf Elliptical Galaxies in the Sculptor Group}
\author{Helmut Jerjen}
\affil{Mount Stromlo \& Siding Spring Observatories, Private Bag, 
Weston Creek P.O., ACT 2611, Canberra, Australia\\
and\\
Astronomical Institute of the University of Basel,
Venusstrasse 7, CH-4102 Binningen, Switzerland}
\authoremail{jerjen@mso.anu.edu.au}
\author{Ken C.\ Freeman}
\affil{Mount Stromlo \& Siding Spring Observatories, Private Bag, 
Weston Creek P.O., ACT 2611, Canberra, Australia}
\authoremail{kcf@mso.anu.edu.au}
\and
\author{Bruno Binggeli}
\affil{Astronomical Institute of the University of Basel,
Venusstrasse 7, CH-4102 Binningen, Switzerland}
\authoremail{binggeli@astro.unibas.ch}
\slugcomment{Accepted by The Astronomical Journal, December 1998 issue}
\lefthead{Jerjen, Freeman \& Binggeli}
\righthead{SBF Distances to Dwarf Elliptical Galaxies}

\begin{abstract} 
As part of an ongoing search for dwarf elliptical galaxies (dE) in the 
vicinity of the Local Group (Jerjen et al.~1998a), we acquired deep $B$ 
and $R$-band images for five dE candidates identified on morphological 
criteria in the Sculptor (Scl) group region. We carried out a surface 
brightness fluctuation (SBF) analysis on the $R$-band images to measure the 
apparent fluctuation magnitude $\bar{m}_R$ for each dE. Using predictions 
from stellar population synthesis models (Worthey 1994) giving $\bar{M}_R$ 
values in the narrow range between $-1.17$ to $-1.13$, the galaxy distances 
were determined. All of these dE candidates turned out to be satellites of Scl 
group major members. A redshift measurement of the dE candidate ESO294-010 
yielded an independent confirmation of its group membership: the 
[\ion{O}{3}] and H$_\alpha$ emission lines from a small \ion{H}{2} region 
gave a heliocentric velocity of 117($\pm5$) \kms, in close agreement with the 
velocity of its parent galaxy NGC 55 (v$_\odot=125$ \kms). The precision of the 
SBF distances (5 to 10\%) contributes to delineating the cigar-like distribution 
of the Scl group members, which extend over distances from 1.7 to 4.4 Mpc and 
are concentrated in three, possibly four subclumps. The Hubble diagram for nine 
Scl galaxies, including two of our dEs, exhibits a tight linear velocity--distance 
relation with a steep slope of 119 \kms\ Mpc$^{-1}$. The results indicate that 
gravitational interaction among the Scl group members plays only a minor role 
in the dynamics of the group. However, the Hubble flow of the entire system 
appears strongly disturbed by the large masses of our Galaxy and M31 leading to 
the observed shearing motion. From the distances and velocities of 49 galaxies 
located in the Local Group and towards the Scl group, we illustrate the continuity of 
the galaxy distribution which strongly supports the view that the two groups form a 
single supergalactic structure.   
\end{abstract}

\section{INTRODUCTION}

It is notoriously difficult to determine distances to dwarf elliptical
galaxies (hereafter dEs, subsuming ``dwarf spheroidals'', cf.~Ferguson
\& Binggeli 1994). Their low gas content rules out HI 21\,cm-line
observations, and their low surface brightness makes optical
spectroscopy feasible only for the very brightest members of the class. 
Most of the available redshifts of dEs are for galaxies in the Virgo
cluster (Binggeli et al.~1993) and the Centaurus cluster (Stein et
al.~1997), and reflect peculiar velocities rather than individual
distances within the cluster. But in clusters there is no real need for
individual distances: dEs are so abundant and strongly clustered (see,
e.g., Binggeli et al.~1987 for Virgo, or Thompson \& Gregory 1993 for
Coma) that they must lie near the mean distance of the cluster.

In poor groups and the field, where the distribution of dEs is sparse, 
individual distance information is absolutely indispensable. (We recall
that true ``field'' (i.e. isolated) dEs are apparently very rare:
cf.~Binggeli et al.~1990). Here the only way to locate a diffuse, 
dwarf-like system and thus to unveil its physical nature is to 
estimate its distance.

For nearby dE candidates, it is possible to resolve the stellar population 
and derive the distance from a deep colour-magnitude 
diagram (CMD, e.g.~Da\,Costa et al.~1996, Smecker-Hane et al.~1996, 
Stetson 1997), or from the tip of the red giant branch (TRGB, Lee et 
al.~1993, Caldwell et al.~1998). These methods are costly and time consuming.
With HST, the CMD method is feasible out to $D \approx$ 4 Mpc, and the 
TRGB method out to $D \approx$ 12 Mpc. Alternatively, the surface 
brightness-magnitude relation of dEs (e.g.~Binggeli \& Cameron 1993, 
Jerjen \& Binggeli 1997) can be used to estimate the distance of a dE out 
to perhaps 50 Mpc. However, the uncertainty (1$\sigma$) in the distance 
modulus for a single galaxy is large, typically $\pm$ 0.7 mag, corresponding
to a 40\% distance error.  Young \& Currie (1995) claimed that this error can 
be reduced to $\pm$ 0.4 mag (comparable to the accuracy of the Tully-Fisher 
method) with a similar relation based on the shape of the dE luminosity profile, 
but this claim appears to be incorrect (Binggeli \& Jerjen 1998).

Here we show that the well-known surface brightness fluctuation (SBF) method
can be used to determine distances to dwarf ellipticals in the 
intermediate distance range of 1--10 Mpc (and possibly beyond) with an 
accuracy of a few percent. The SBF method was introduced by Tonry \& 
Schneider (1988, hereafter TS88) to measure distances to bright elliptical 
(E) galaxies. The method is based on the discrete sampling of a galaxy image 
with a CCD detector and the analysis of the resulting pixel-to-pixel variance 
caused by unresolved stars. For a detailed description of the method the reader 
is referred to TS88 and Jacoby et al.(1992). Application of the SBF method was 
subsequently extended to bulges of spiral galaxies (Tonry 1991, Luppino \& Tonry 
1993) and to globular clusters (Ajhar \& Tonry 1993). A massive SBF observing 
programme is in progress to improve the cosmic distance scale (Tonry et al.~1997). 

The first, and up to now only, application of the method to {\em dwarf}\/ 
ellipticals is due to Bothun et al.~(1991) who observed the fluctuations in 
very large, nearly flat, low-surface brightness galaxies (a special type of 
dEs) in the Fornax cluster of galaxies. We show here that the SBF method is 
particularly well suited for dEs at a distance of a few Mpc -- slightly beyond the 
distance where the galaxies would be resolved into stars.  

In this paper we present a pilot SBF analysis for five dwarf ellipticals in 
the nearby Sculptor (Scl) group, also known as the South Polar group. 
These galaxies have recently been identified in an extensive survey of the 
southern Scl and Centaurus\,A (Cen\,A) groups for faint dE members 
(Jerjen et al.~1998a: JBF98a). Special care has been taken
with the calibration of the fluctuation magnitude $\bar{M}_R$, because
the stellar population of dwarf ellipticals is quite distinct from that of
the brighter ellipticals. In fact, even among the local dwarf spheroidals there 
is a wide range of star formation histories. Nevertheless, 
we find that $\bar{M}_R$ is sufficiently robust against such variations, 
at least for stellar systems dominated by old populations, and
propose a first estimate of $\bar{M}_R$ for dEs. 

We will show that the SBF distances derived for the five dEs (between 1.7 and 
4.4 Mpc), unambiguously identify these objects as Sculptor group members. 
Moreover, the dwarfs neatly fall into place with the known substructure of the 
group. We demonstrate the great potential of the SBF method applied to dEs as 
a tool to map the supergalactic structure in the local volume (out to the Virgo 
cluster). With our enlarged Scl sample, we re-analyse the 3D distribution of 
these galaxies and confirm earlier findings (Tully \& Fisher 1987, Binggeli 1989) 
that the Sculptor complex has a prolate structure and is part of a much larger 
cloud of galaxies which includes the Local Group (LG). The SBF results 
for the dE sample from our Cen\,A group survey will be published 
elsewhere (Jerjen et al.~1998b: JFB98b).

\section{SAMPLE}  

Following C\^ot\'e et al.'s~(1997, hereafter C97) successful search of the Scl 
group region for faint new dwarf irregular members, we conducted a similar 
visual  search for faint dE candidates using the same plate material. Seven very 
diffuse objects were identified which had dE-like morphology and which were 
thus suspected to be dwarf spheroidal members of the Scl group 
(for the details of this survey, see JBF98a). Subsequent CCD imaging in the 
$B$ and $R$ bands showed two of these objects to be irregulars or 
background galaxies. The five remaining galaxies of early-type morphology 
are the subject of the present SBF analysis. They are listed in Table~\ref{tbl-1}  
with their basic photometric properties from JBF98a. 

\begin{deluxetable}{llcccccccc}
\scriptsize
\tablecaption{Galaxy properties \label{tbl-1}}
\tablewidth{0pt}
\tablehead{
\colhead{Galaxy} &\colhead{Type} & \colhead{RA} & \colhead{DEC}  &\colhead{$R_T$} & \colhead{$r_{eff,R}$} & 
\colhead{$\langle\mu\rangle_{eff,R}$} &\colhead{$(B-R)_T$} & \colhead{A$_B$} & \colhead{$(B-R)_T^0$} \\
\colhead{}       & \colhead{}    &\colhead{(2000)}& \colhead{(2000)}      & \colhead{mag}          & \colhead{arcsec} &
\colhead{mag\,arcsec$^{-2}$}& \colhead{mag}     & \colhead{mag}   & \colhead{mag}      \\
\colhead{(1)}    & \colhead{(2)}              & \colhead{(3)}   & \colhead{(4)}      & \colhead{(5)}          & \colhead{(6)} &
\colhead{(7)}    & \colhead{(8)}              & \colhead{(9)}   & \colhead{(10)}      
}     
\startdata  
NGC 59         &dS0   &00 15 25.1 & $-$21 26 38 & 11.90 & 25.9       &  21.0          & 1.07         &  0.09     &  1.04     \nl 
Scl-dE1 (SC22\tablenotemark{1}\,\,) &dE &00 23 51.7 & $-$24 42 18 & 16.94 & 19.3 &  25.4 & 0.79      &  0.06     &  0.77     \nl 
ESO294-010     &dS0/Im&00 26 33.4 & $-$41 51 19 & 14.36 & 20.3       &  22.9          & 1.17         &  0.03     &  1.16     \nl 
ESO540-030     &dE/Im &00 49 21.1 & $-$18 04 34 & 15.54 & 21.5       &  24.2          & 0.83         &  0.10     &  0.79     \nl 
ESO540-032     &dE/Im &00 50 24.5 & $-$19 54 23 & 15.36 & 25.0       &  24.4          & 1.08         &  0.09     &  1.05     \nl 
\enddata
\tablenotetext{1}{Name from C\^ot\'e et al.~(1997)}
\end{deluxetable}

The photometric parameters span a wide range in total $R$ magnitude 
($11.9<R_T<17.0$, column 5), effective radius which contains half of the total 
light ($19''<r_{eff,R}<26''$, column 6), and mean surface brightness within the 
effective radius ($21.0< \langle\mu\rangle_{eff,R}<25.4$, column 7). To get the 
extinction-corrected colour $(B-R)_T^0$ (column 10) the foreground reddening 
E($B-V$) was determined from the dust maps of Schlegel et al.~(1998) and converted 
with the ratio $A_B:A_R:\mbox{E}(B-V)=4.315:2.673:1$. Further details about 
the photometric data can be found in JBF98a.

\begin{figure*}
\centering\leavevmode
\epsfxsize=6.5cm
\epsfbox{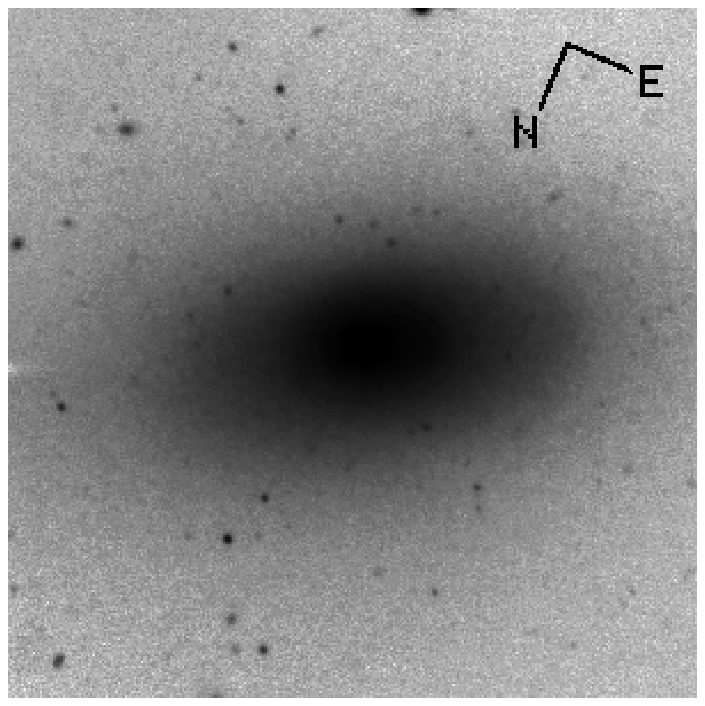}
\hspace{0.2mm}
\epsfxsize=6.5cm
\epsfbox{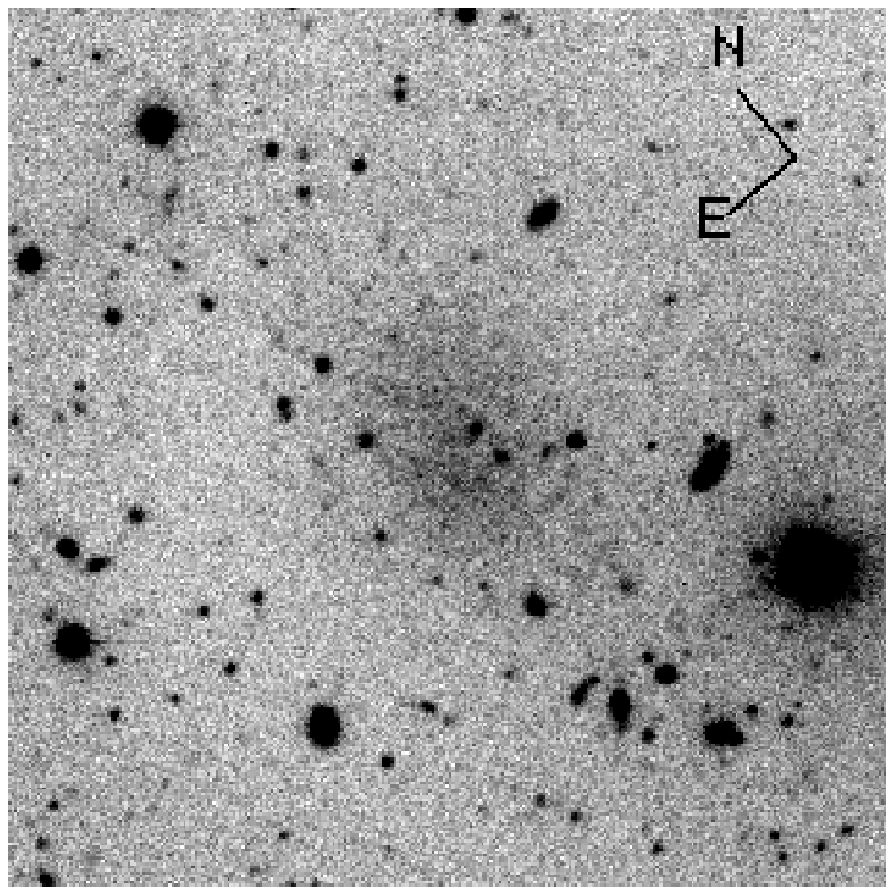}\\
\centering\leavevmode
\epsfxsize=6.5cm
\epsfbox{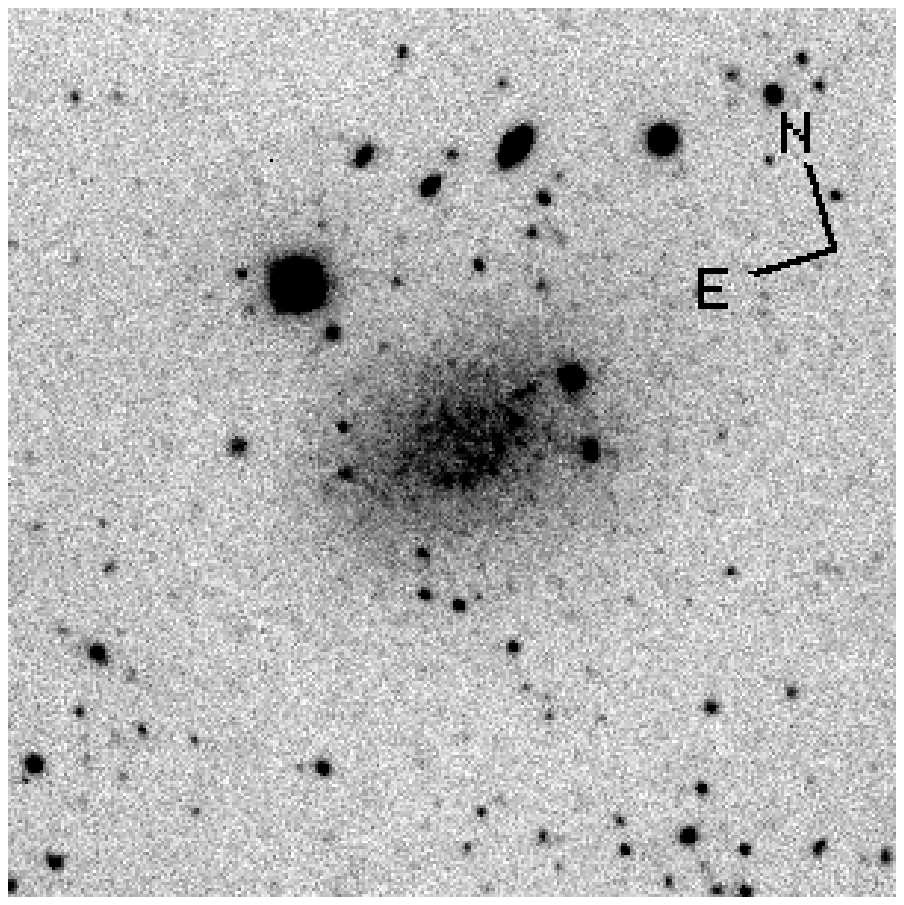}
\hspace{0.2mm}
\epsfxsize=6.5cm
\epsfbox{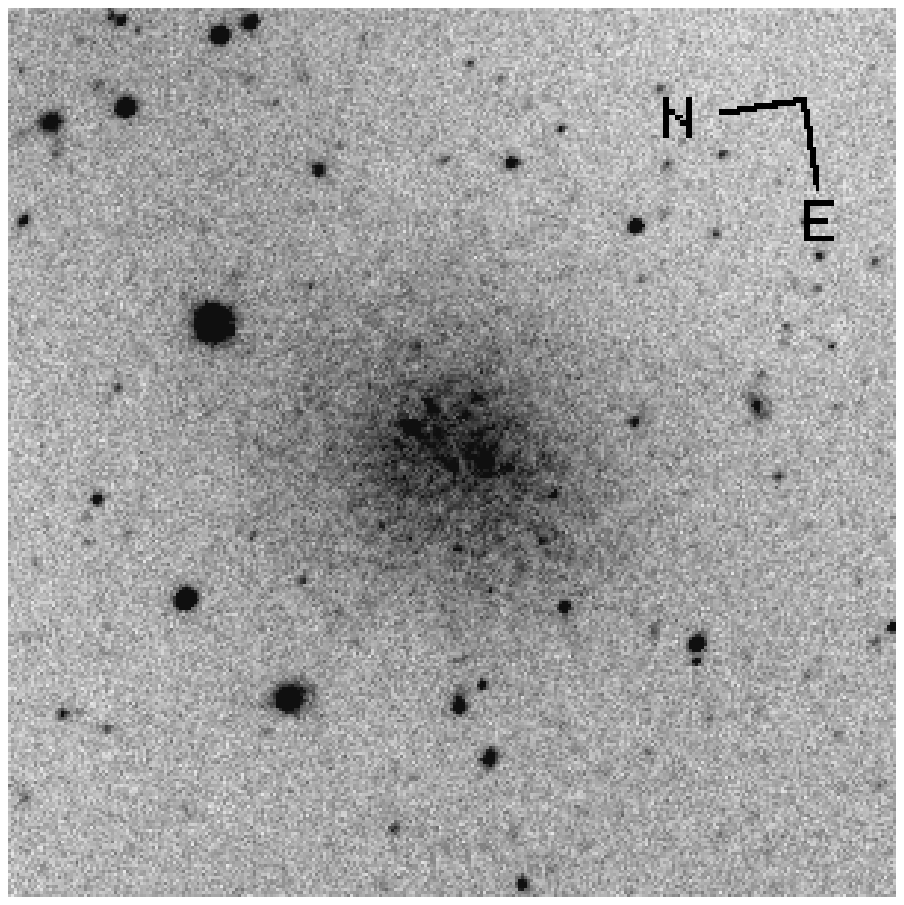}\\
\caption{$R$-band images of four Sculptor group dwarfs made from
the combination of 4 -- 5 exposures with a total integration time
between 1800 and 3000 seconds. The areas depicted are 3 arcmin on a 
side. From left to right and top to bottom the galaxies are: NGC 59, 
Scl-dE1, ESO540-030, and ESO540-032. \label{fig1}}
\end{figure*}

\begin{figure*}
\centering\leavevmode
\epsfxsize=6.5cm
\epsfbox{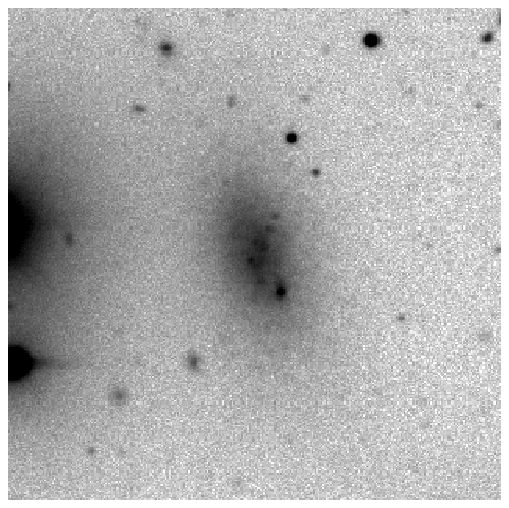}
\hspace{0.2mm}
\epsfxsize=6.5cm
\epsfbox{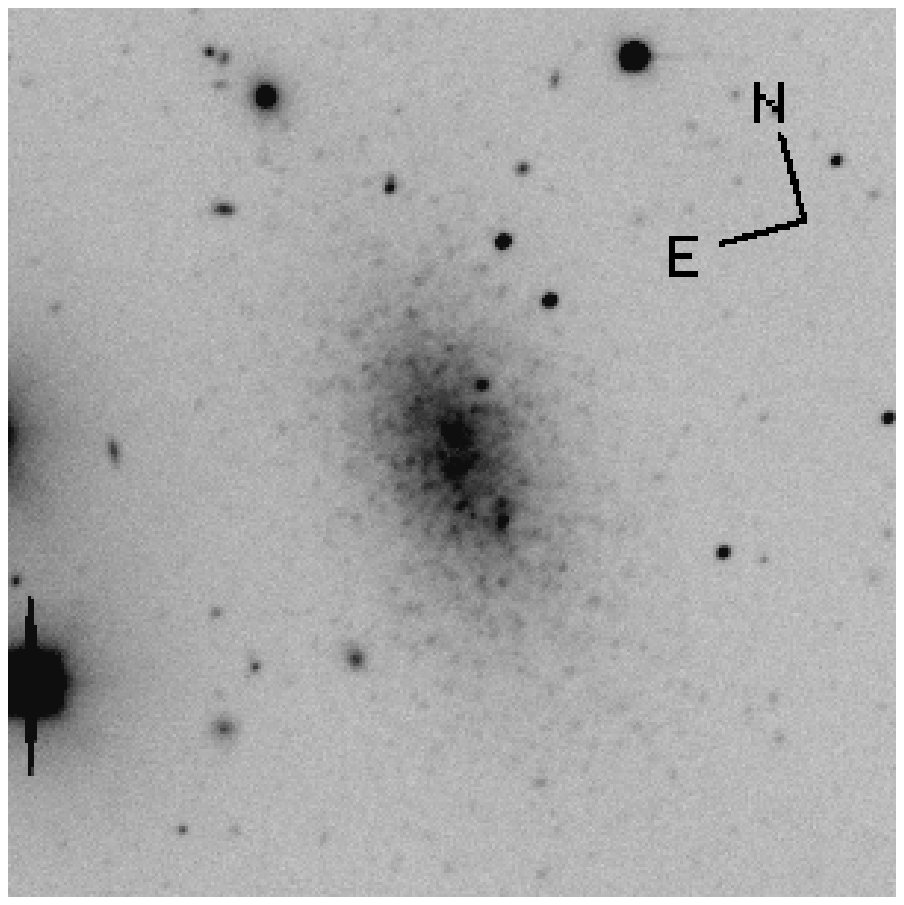}
\caption{$B$ and $R$-band images of ESO294-010 made from the combination of 
5 exposures with total integration times of 2500 and 2250 seconds, 
respectively. The angular frame size is 3 arcmin. The possible \ion{H}{2} 
region is the bright feature 18 arcsec south of the centre of the galaxy. 
\label{fig2}}
\end{figure*}

Images of the dwarfs are shown in Figs.~\ref{fig1} and \ref{fig2}, and 
the classifications given in column 2 of Table~\ref{tbl-1}. There is some 
morphological variety among early-type dwarfs (cf.~the atlas of Virgo cluster 
dwarfs by Sandage \& Binggeli 1984). At the bright end of the luminosity 
function, there is a variant of the pure dE which
was named dS0, because it is often distinguished by a S0-like, two-component 
structure (see also Binggeli \& Cameron 1991). NGC 59 is clearly of this type,
ESO294-010 arguably so. At fainter magnitudes, dEs (dwarf spheroidals)
are often hard to distinguish from smooth irregulars of type Im\,V. In fact,
there may be a true evolutionary transition between gas-rich irregulars
and gas-poor dEs (e.g. Ferguson \& Binggeli 1994). Dwarfs with a mixed
morphology (appearing too smooth for a plain Im, but too lumpy for a pure dE)
were called ``intermediate'' or ``ambiguous'' and classified dE/Im
(Sandage \& Binggeli 1984). A well-known local example is the Phoenix dwarf
system (Van de\,Rydt et al.~1991). 

Three of our objects seem to fall in this class and are classified 
accordingly. ESO540-030 and ESO540-032 show a sprinkle of resolved stars in 
their central regions. These stars may mark the presence of a small population of
young stars. ESO294-010, on the other hand, appears slightly lumpy in the 
centre (see Fig.\ref{fig2}). Subsequent spectroscopy (in Sec.6) showed emission 
lines, so this is a clear example of a mixed type. But even the smoothest dEs 
probably contain small amounts of intermediate-age or young stars; this we know 
from the local dwarfs (cf.~Sec.5). Note also that the colours of our objects
(with $0.75<B - R<1.04$, cf.~column 10 of Table~\ref{tbl-1}) are rather blue for dEs,
whose $B - R$ values range from 0.5 to 2.3 (e.g., Evans et al.~1990).
So the morphological deviations from a pure dE are simply the manifestation of 
an underlying variety in the (unresolved) stellar populations. One of the principal 
results of the present paper is that this variety does not cause difficulties
for the application of the SBF method to dEs, because the fluctuation magnitude 
$\bar{M}_R$ turns out to be rather insensitive to variations in the stellar contents 
within the observational constraints (Sec.5). In the following we will use the term 
``dE'' in a loose way, including all of the morphological variants discussed above.
  
\section{Imaging}

The galaxies were observed in two runs, on September 11--15, 
1996, and August 28--30, 1997, using the imager at the Nasmyth B focus 
of the 2.3m ANU telescope at Siding Spring Observatory. The detector was a 
Tek 1k$\times$1k thinned CCD with a pixel size of 24$\mu$m, yielding a 
scale of 0.6$''$ pixel$^{-1}$ and a 6.7$'$ diameter circle field of view. The 
CCD gain was 1 e$^{-}$/ADU and the readout noise was 7 e$^{-}$. We  
observed in the Kron-Cousins $R$-band, because our application of the SBF 
technique differs from previous studies which used $I$ or $K$; we are working 
on small, low-surface brightness objects so our data are 
photon-limited by the sky rather than by the galaxy itself. The sky is 
significantly fainter in $R$ than in $I$ relative to the RGB stars 
which produce the surface brightness fluctuations [$(R-I)_{sky}$=1.5 versus 
$(R-I)_{RGB}$=0.8]. Another advantage of the $R$ filter is to avoid fringing 
which occur with thinned CCDs beyond 7000\AA. 

Weather conditions were photometric and the seeing measured between $1.2-1.5''$. 
For each of our galaxies a series of 4 to 5 exposures were taken each randomly 
offset by $\sim10''$ and of 450 to 600 sec duration, giving a total integration time 
of 1800 to 3000 sec. Flat fields were obtained every night from exposures of the 
morning and evening sky. Standard stars of Graham (1982) were observed throughout
each night for the photometric calibration. 

Processing of the CCD frames with IRAF was carried out in the usual manner. 
The overscan region provided the bias level, which was subtracted from the 
frame. Next, each science frame was flattened with a master twilight-sky flat,
constructed from a median of 5--9 sky flats taken during the same night. All 
reduced frames were flat to 0.1\%. The sky was modelled by fitting a plane to 
selected star-free areas well away from the galaxy. While sky determinations can be 
quite difficult for giant ellipticals due to their extended halos and galaxy sizes 
comparable to the field of view, this is simple in our case. Finally, the sky 
subtracted images of a galaxy were registered and combined to a master frame in order 
to increase the signal-to-noise ratio and to remove artifacts, bad pixels, and cosmic 
rays. In Fig.\ref{fig1} we show the processed $R$-band images for four of our 
galaxies. $B$ and $R$-band images of ESO294-010 are shown in Fig.\ref{fig2}. 

\section{SBF ANALYSIS}

To prepare the master frames for the SBF analysis we essentially followed  
the procedure described in TS88. First, we employed DAOPHOT (Stetson 1987) 
routines to identify point sources which would disturb a 2D-fit of the mean 
galaxy light distribution. Because of the small size of the galaxies
and the high galactic latitude, there is only little contamination by 
foreground stars and background galaxies. Globular clusters 
are also a minor problem here, because globular clusters are rare in 
faint dwarf galaxies. However, as mentioned in Sec.2, two 
of our galaxies, classified as dE/Im, are contaminated with a few bright 
stars of their own. These also had to be removed.

All identified point sources were replaced by a nearby patch of galaxy
of the same surface brightness. Isophotes were computed for the cleaned
galaxy image by fitting ellipses with variable radius, ellipticity, and
position angle. These were used to model the mean galaxy surface
brightness distribution, which was then subtracted from the original
master frame. To normalise the variance amplitude of the stellar
fluctuations across the residual image, we divided the residual image
by the square root of the mean galaxy surface brightness distribution.

For each galaxy the SBF analysis was carried out on two different square 
subimages (field 1 and 2) with size between 30 to 60 pixels which were 
selected within the 27.5 $R$\,mag\,arcsec$^{-2}$ isophote. The overlap of 
the fields was kept as small as possible ($<10$\%) to get independent 
distance measurements. Special care was further taken to choose subimages 
with only few ($<10$) resolved point sources, galaxies, or other features 
DAOPHOT formerly had identified. For example, the \ion{H}{2} region of 
ESO294-010 (see Sec.6 for more details) was excised from the analysis. 
Remaining objects in the subimages were replaced by uncontaminated patches 
(i) randomly selected from the region outside of the subimage area and (ii) 
in the same surface brightness range. This patching process was employed as 
alternative to the masking method described in TS88. The idea was to 
replace the few disturbed image parts with patches carrying the 
fluctuation signal. As artificial periodicities and poor signal-to-noise 
data could be introduced in this way it is crucial to take into account 
criteria (i) and (ii). The number of pixels corrected never exceeded 5\% 
of the total subimage area. Experiments showed that, at this percentage 
level, the patching method is uncritical for the signal we intend to 
measure.

\begin{figure*}
\centering\leavevmode
\epsfxsize=4.8cm
\epsfbox[25 150 570 580]{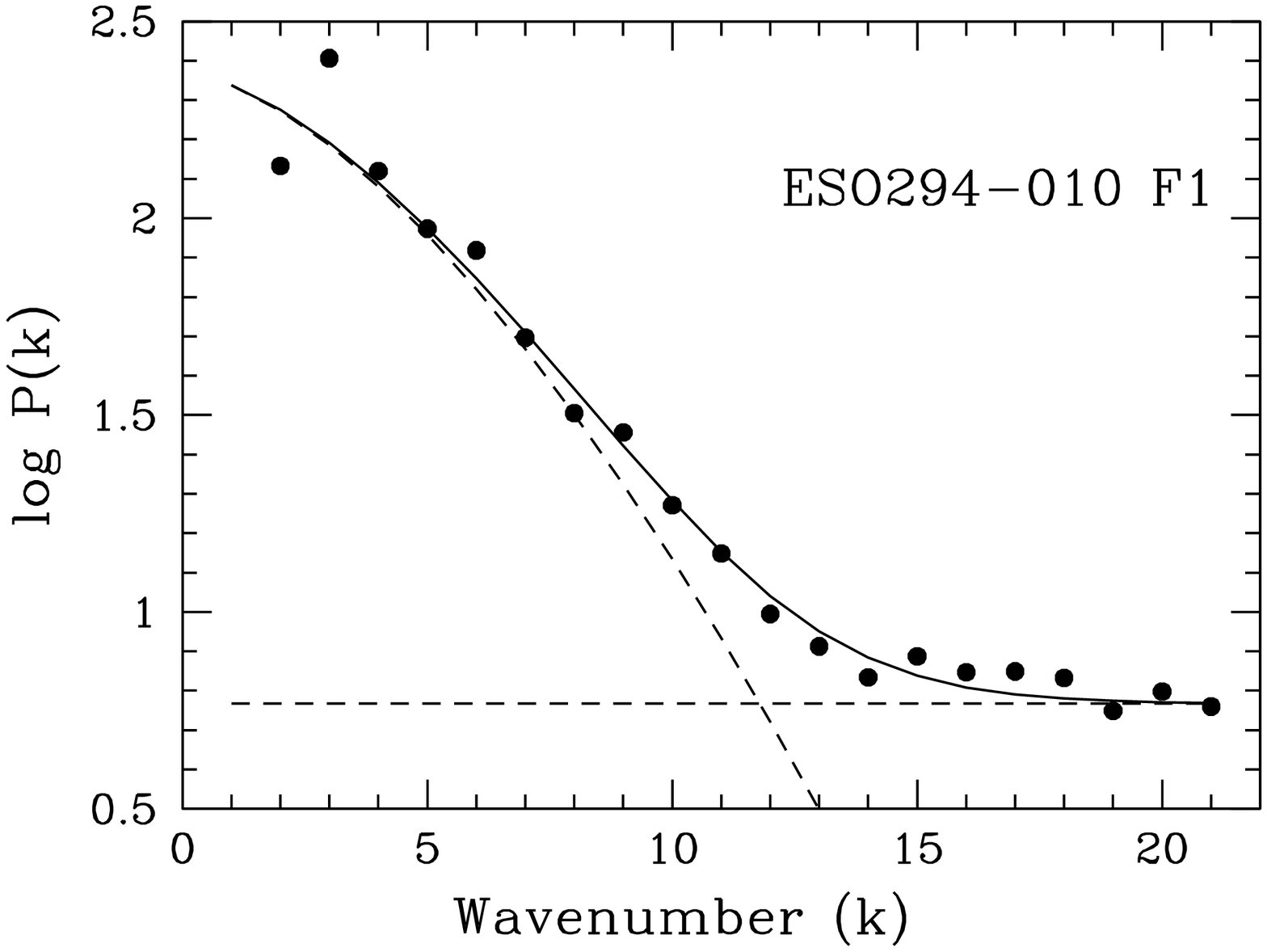}
\hspace{1cm}
\epsfxsize=4.8cm
\epsfbox[25 150 570 580]{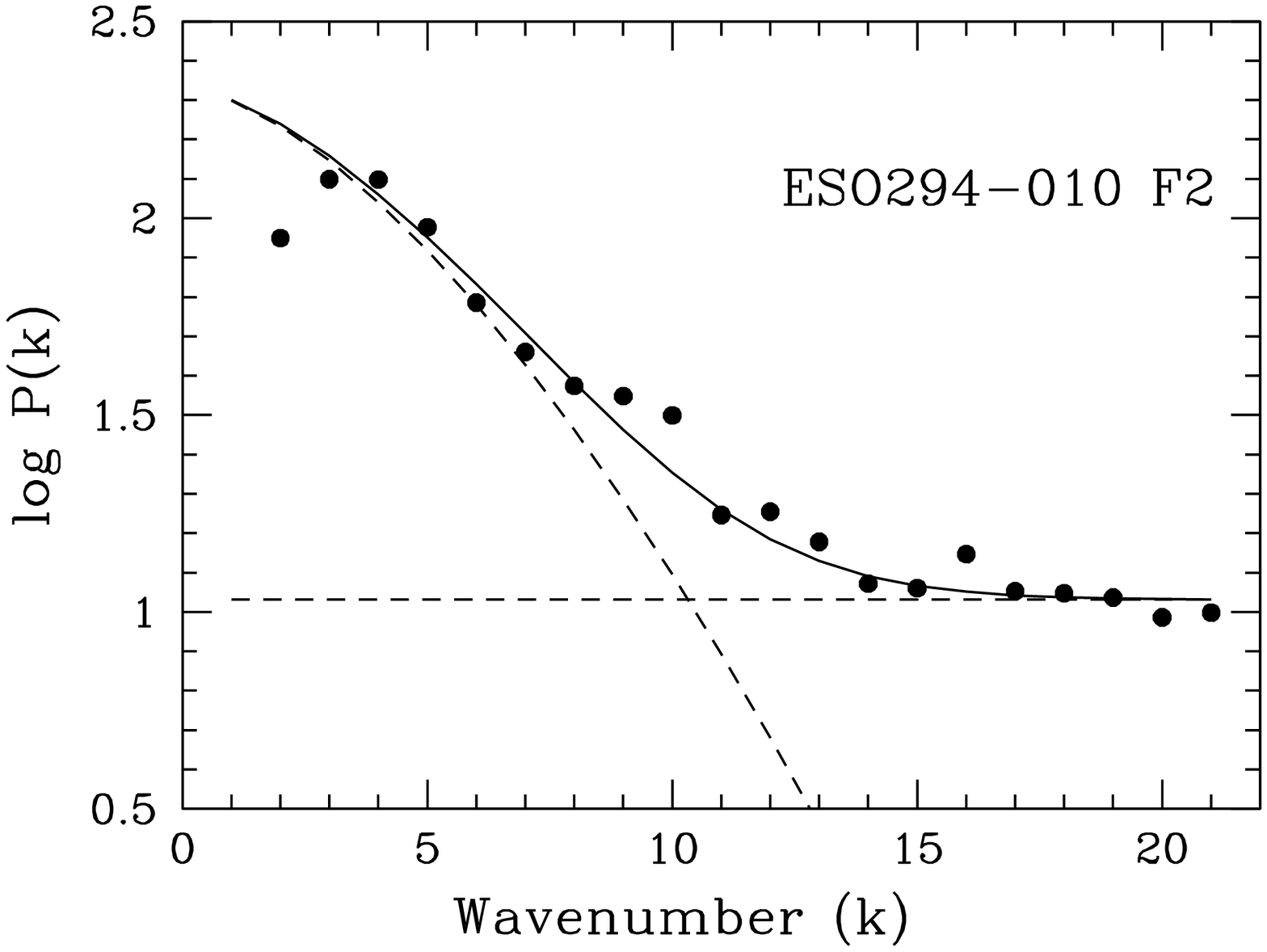}\\
\centering\leavevmode
\epsfxsize=4.8cm
\epsfbox[25 150 570 580]{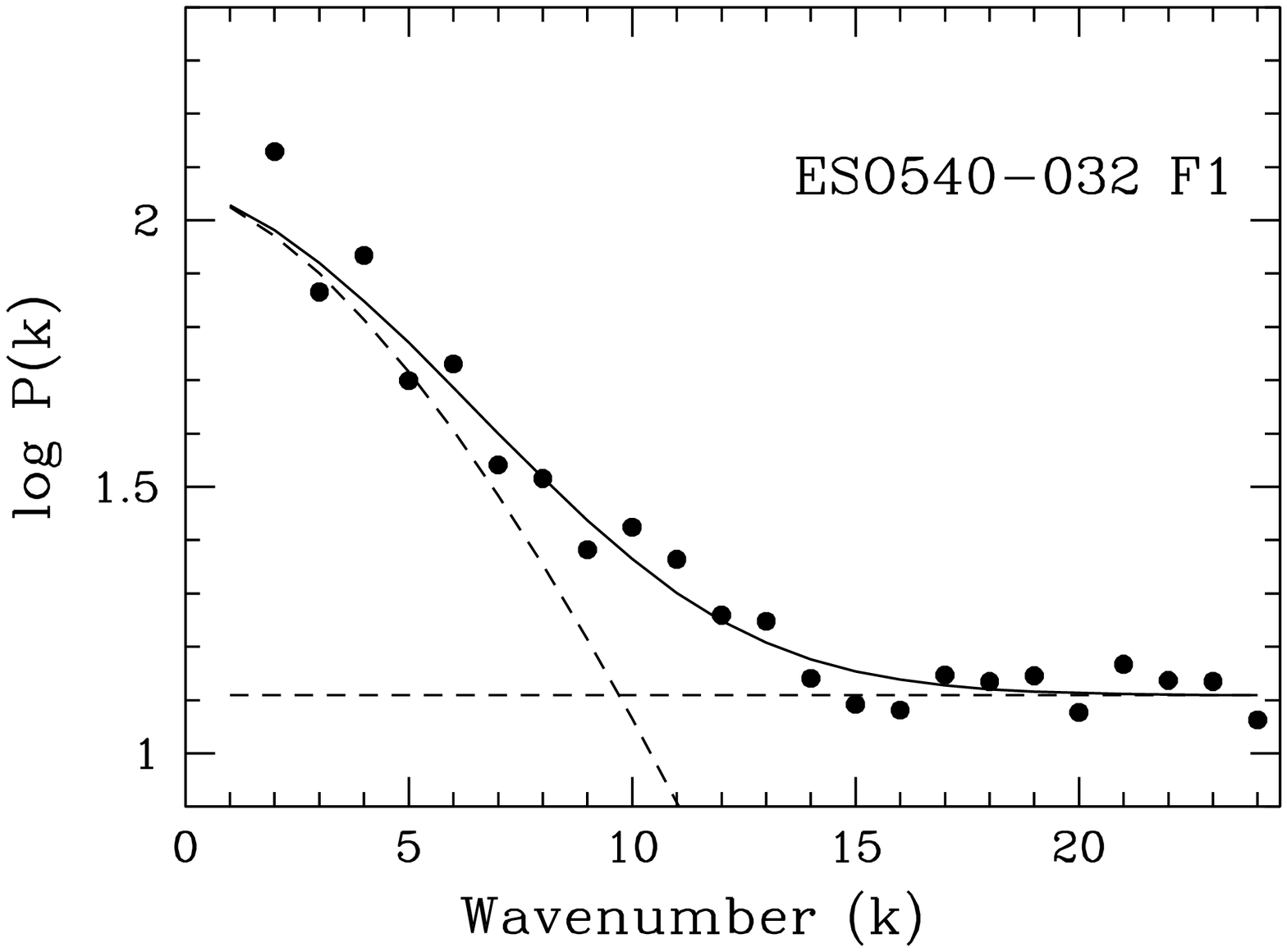}
\hspace{1cm}
\epsfxsize=4.8cm
\epsfbox[25 150 570 580]{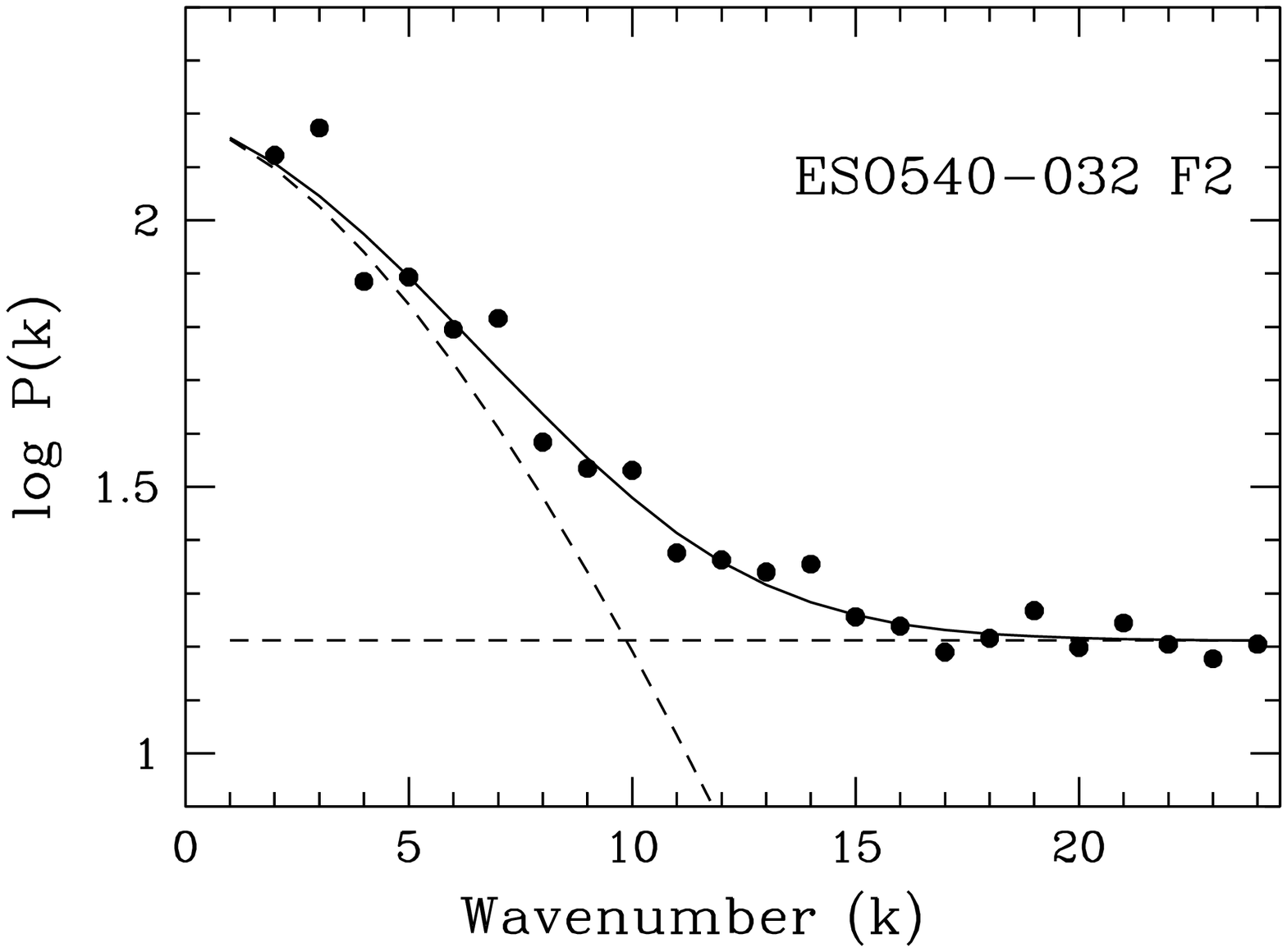}\\
\centering\leavevmode
\epsfxsize=4.8cm
\epsfbox[25 150 570 580]{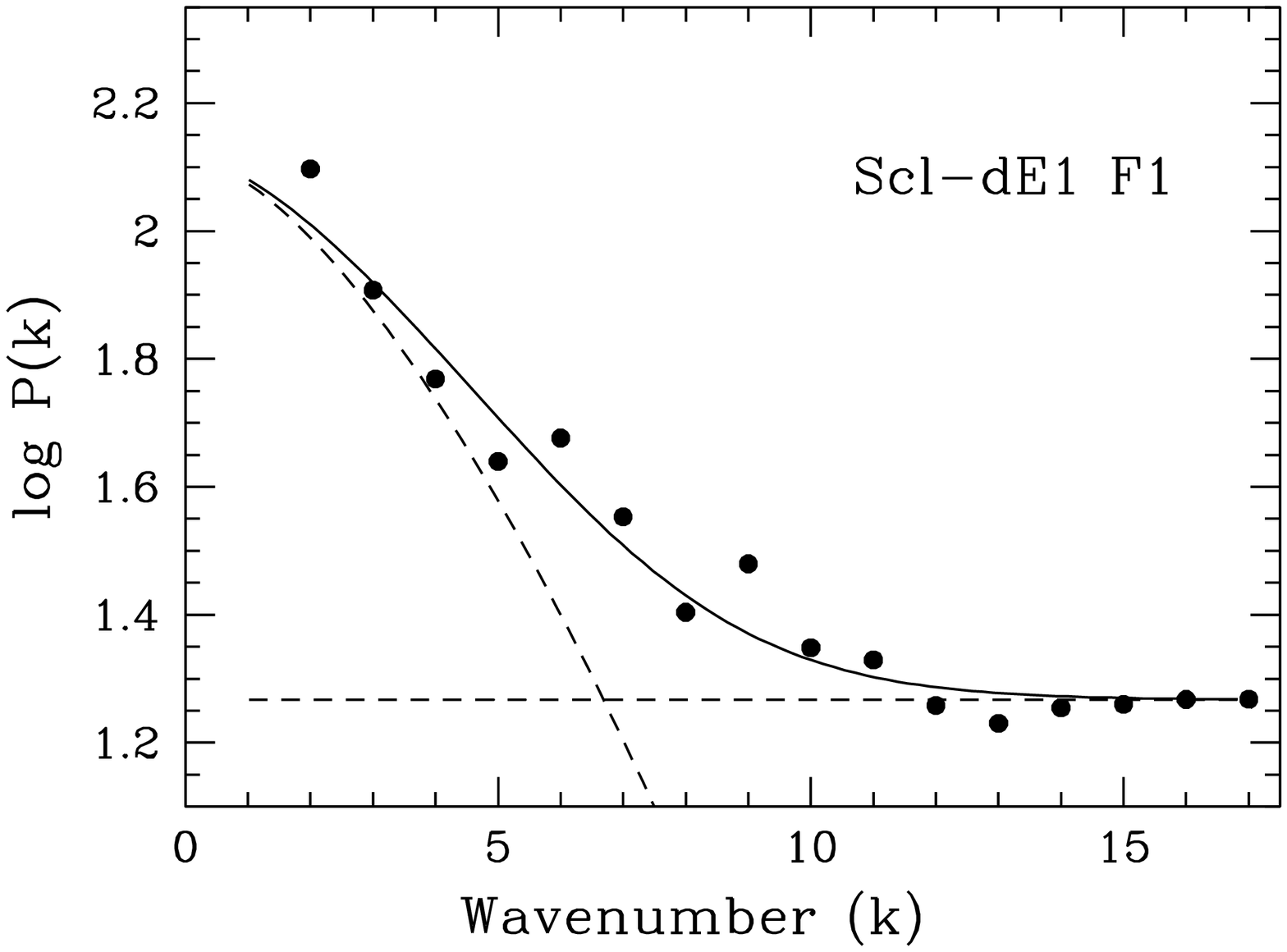}
\hspace{1cm}
\epsfxsize=4.8cm
\epsfbox[25 150 570 580]{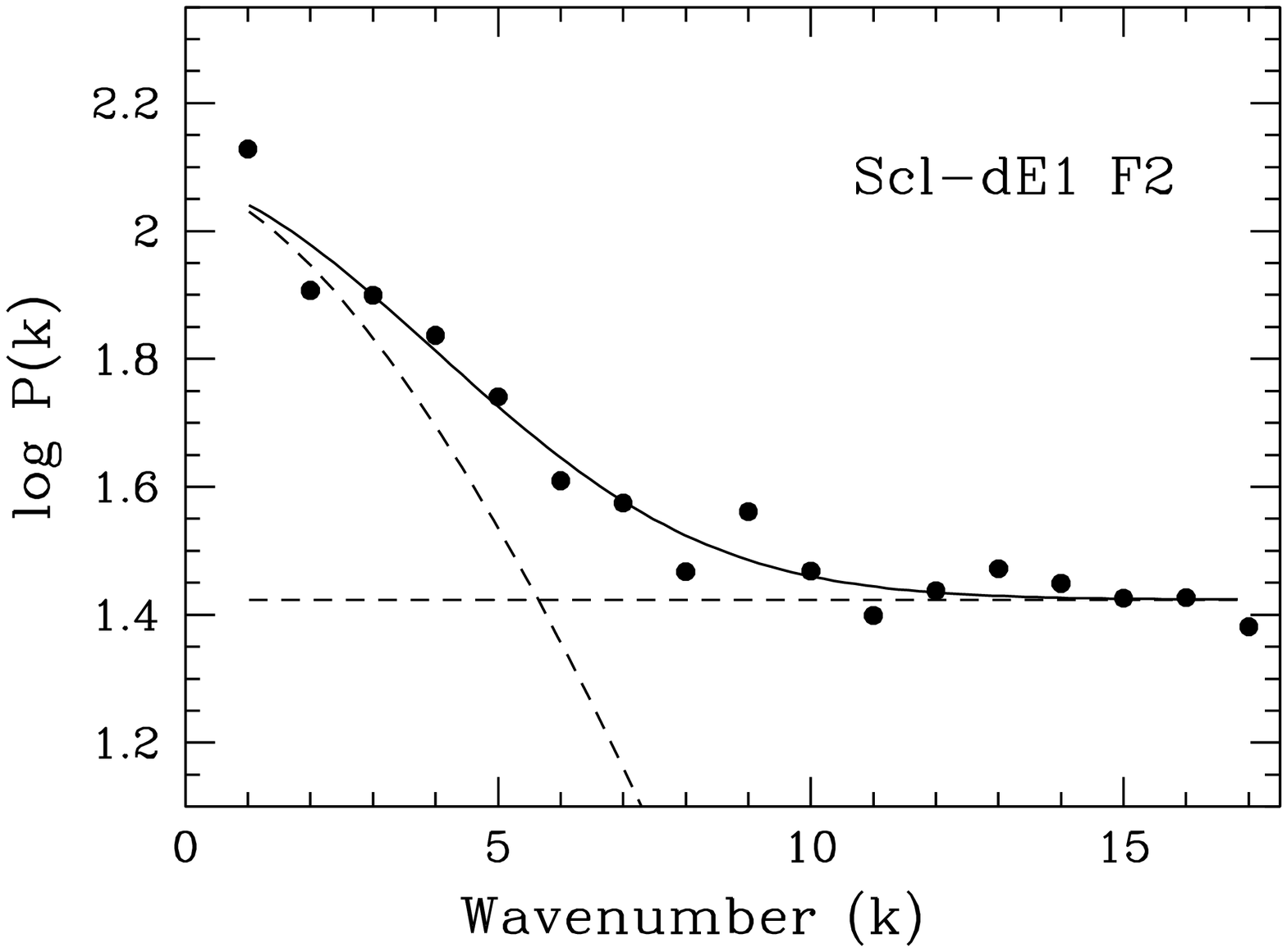}\\
\centering\leavevmode
\epsfxsize=4.8cm
\epsfbox[25 150 570 580]{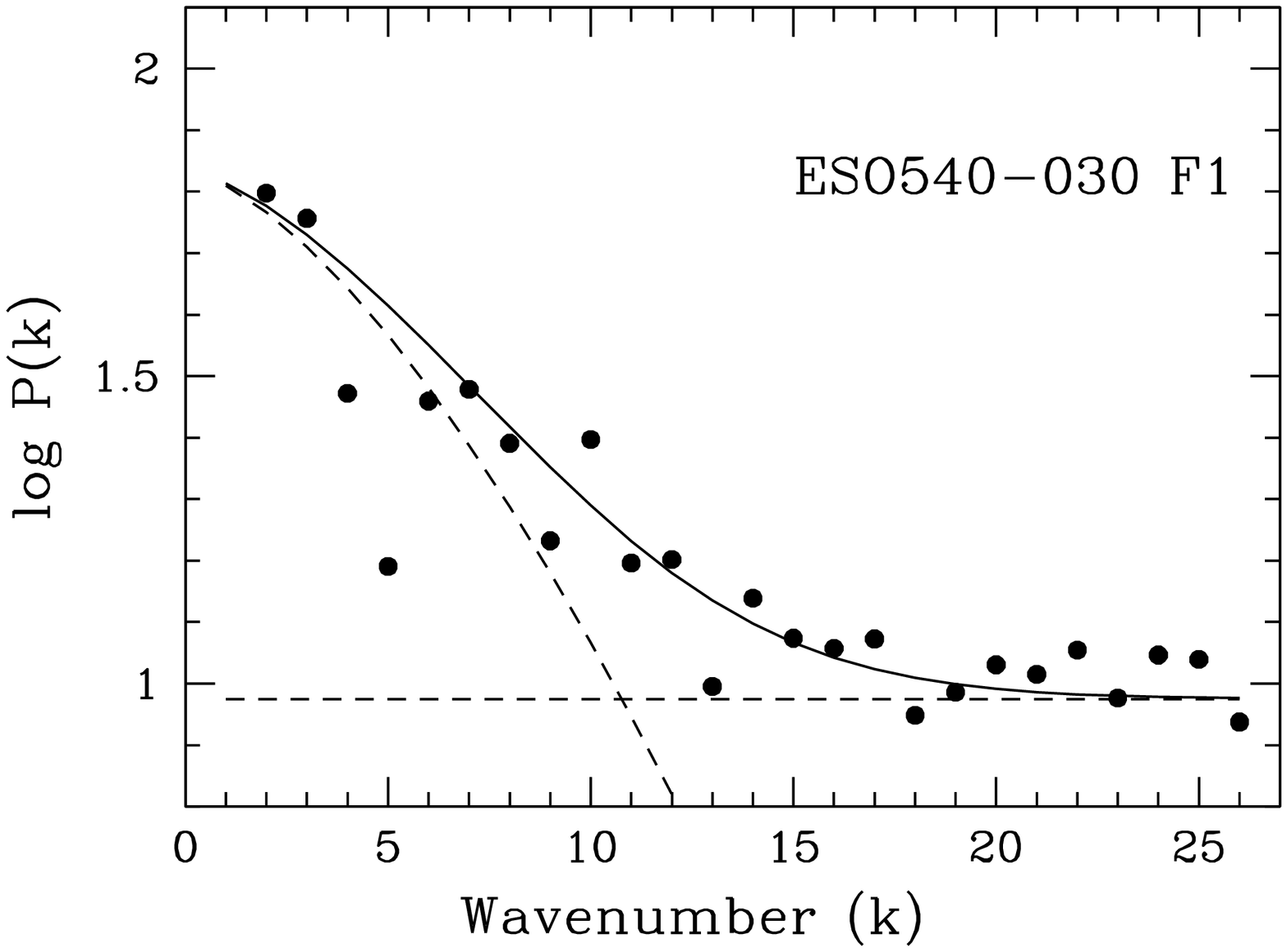}
\hspace{1cm}
\epsfxsize=4.8cm
\epsfbox[25 150 570 580]{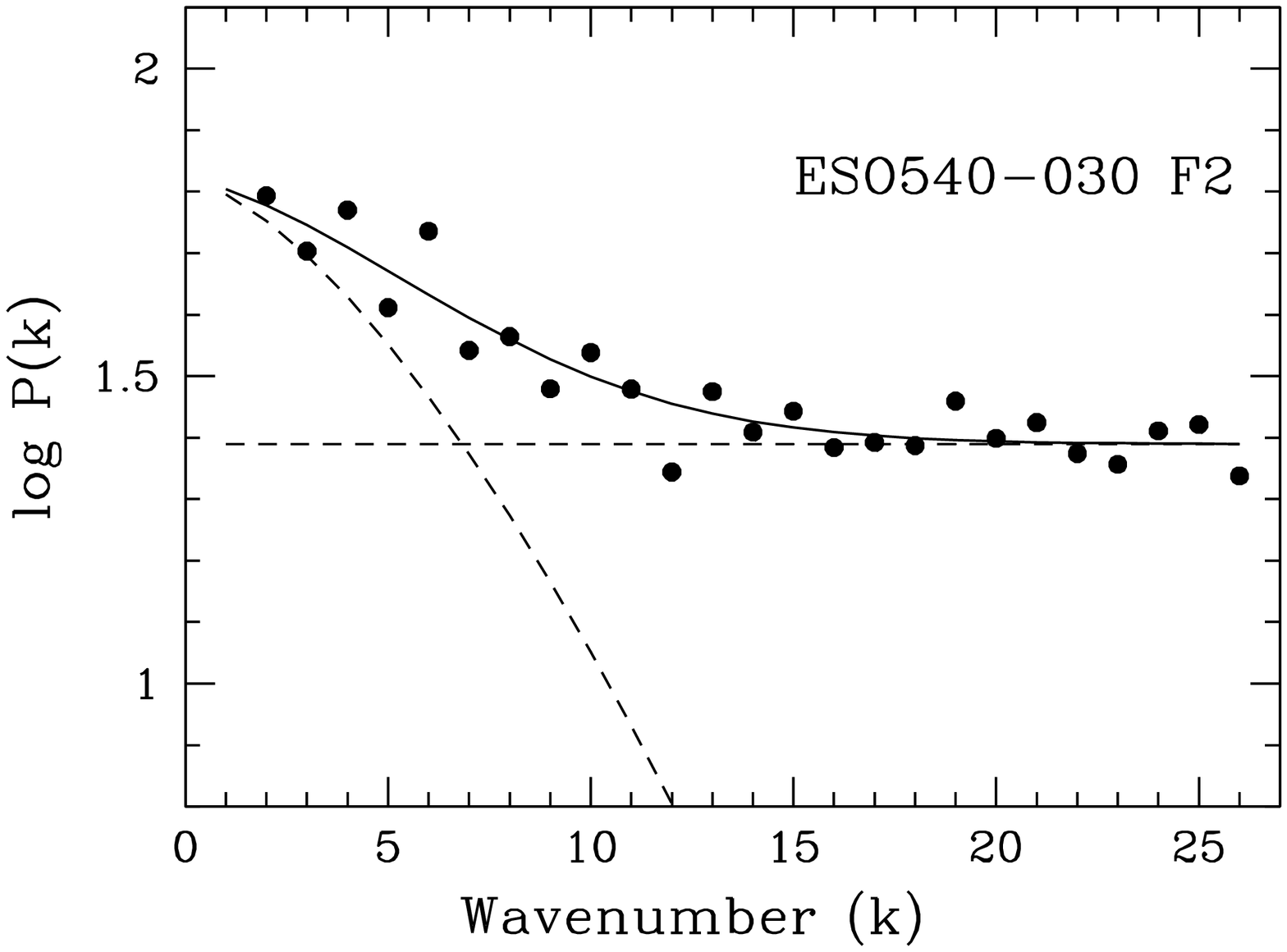}\\
\centering\leavevmode
\epsfxsize=4.8cm
\epsfbox[25 150 570 580]{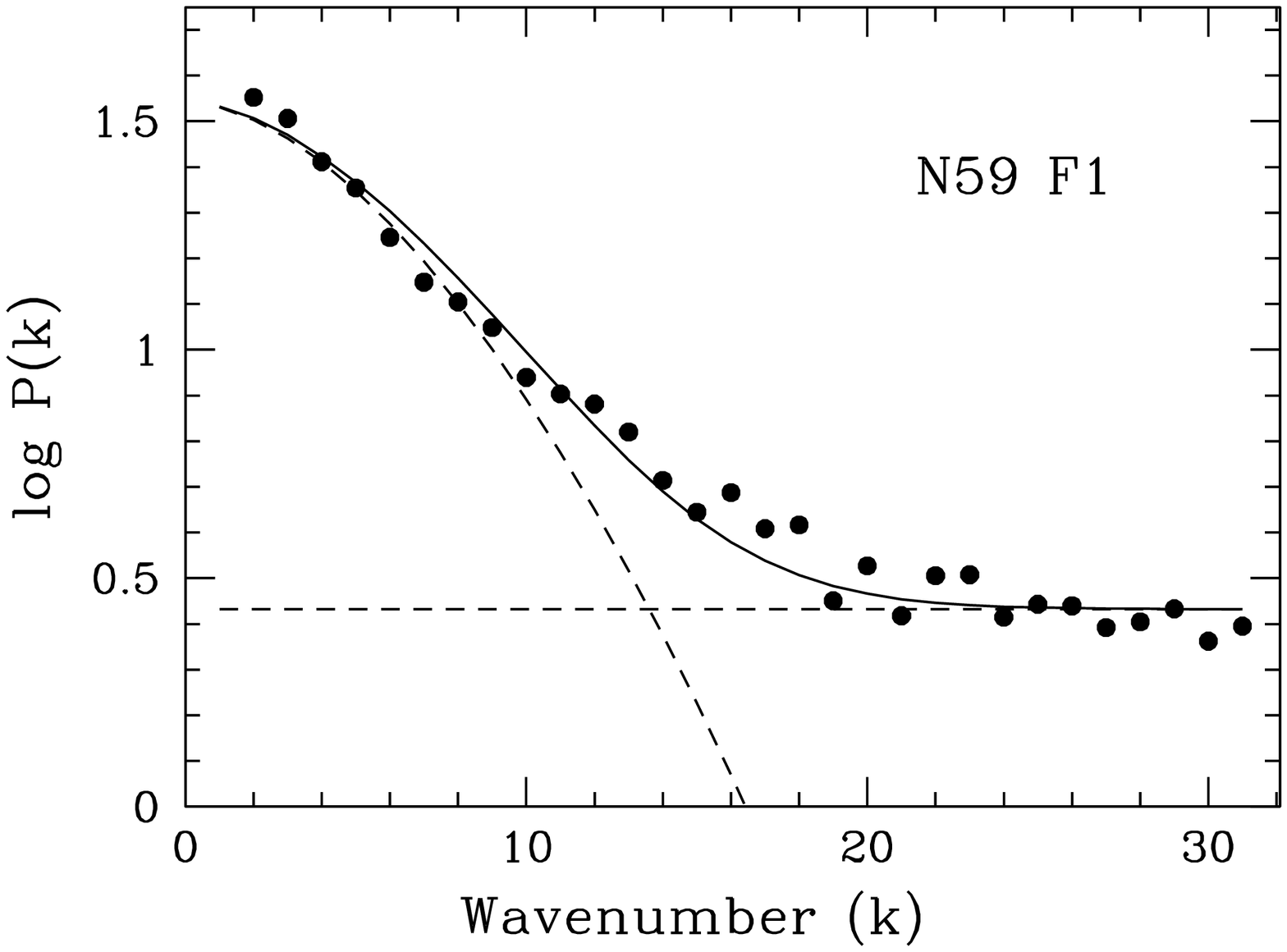}
\hspace{1cm}
\epsfxsize=4.8cm
\epsfbox[25 150 570 580]{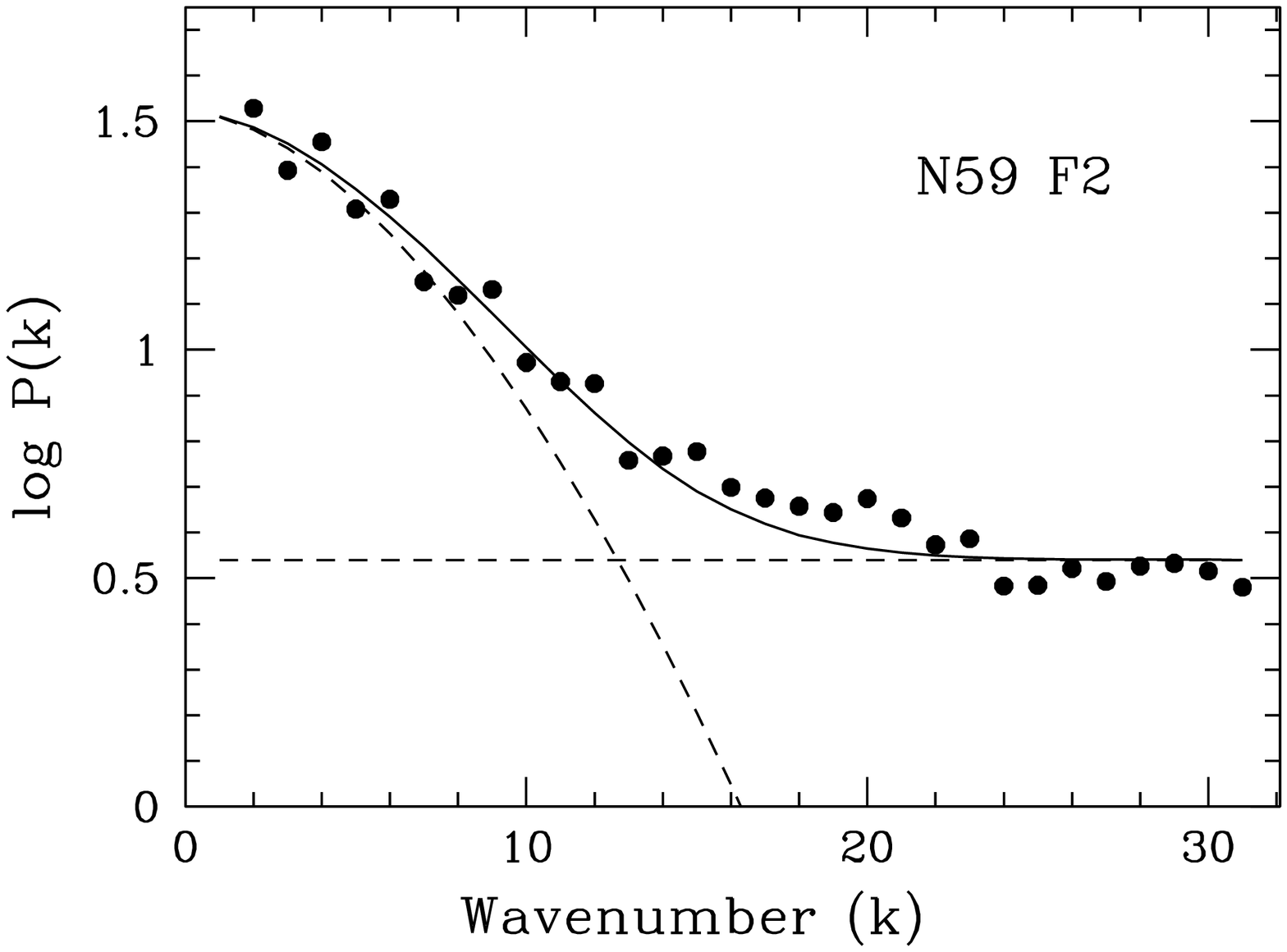}\\
\caption{Power spectra of the Fourier transform for the $R$ subimages 
analysed. There are two fields for every dwarf, denoted F1 and F2.
The spectra are azimuthally averaged. The points are the data. The full lines 
are best-fitting combinations of a PSF-convolved stellar fluctuation component 
and a constant photon noise component, as modelled by Equ.1. The separate 
contributions are shown as broken lines. \label{fig3}}
\end{figure*}

At this stage of the procedure, we have images whose main sources 
of variance are the stellar fluctuations and a photon shot noise that is 
nearly uniform over the image (the CCD readout noise is negligible). To 
disentangle the two components we Fourier-transformed the cleaned subimages and 
analysed their power spectra (PS), which are shown in Fig.~\ref{fig3}. Having 
calculated the PS of the Point Spread Function (PSF) from well isolated stars 
on the master frame, the observed PS of the galaxy was modelled as a weighted 
combination of a PSF-convolved component and a constant white-noise component:
\begin{equation} 
\mbox{PS(galaxy)}=P_0\cdot \mbox{PS(PSF)} + P_1
\end{equation}
The free parameters $P_0$ and $P_1$ (the two weights) were 
determined by a least-squares fit of Equ.1 to the data. The fit was restricted 
to wave numbers $k>4$ to exclude the range which is affected by the galaxy 
subtraction. The result of this decomposition for every galaxy field is shown
in Fig.~\ref{fig3}.

The quantity $P_0$ is directly linked to the apparent fluctuation magnitude 
$\bar{m}_R$ which is the luminosity-weighted average stellar luminosity of the 
stellar population and is approximately the apparent magnitude of a giant 
star in the galaxy which contributes to the SBF signal:   
\begin{equation}
\bar{m}_R=m_1-2.5\cdot\log(P_0/t)\,.
\end{equation}
The system (telescope-filter-detector)-related quantity $m_1$ is 
the magnitude of a star yielding 1 ADU per second on the CCD, and $t$ is
the exposure time of the observation. 

From theory (TS88), we expect $P_1$ to be the variance of the photon shot 
noise $\sigma_{ph}$ divided by the square root of the number $n$ of exposures: 
\begin{equation}
P_1=(1+s/\bar{g})/\sqrt{n}\,,
\end{equation}
where $s$ is the flux from the sky ($\langle \mu_{sky, R}\rangle =20.8$ 
mag\,arcsec$^{-2}$ at new moon) and $\bar{g}$ is the mean surface 
brightness of the galaxy on the subimage (in counts). That Equ.3 is 
approximately fulfilled can easily be verified from Table~\ref{tbl-2} 
where all relevant parameters are listed. 

The sky brightness of individual images varied a few percent due 
to changes in the airglow intensity during the night. Column 6 gives 
the average sky value we measured for each series of images. The indicated 
error corresponds to the uncertainty in the sky flatness of the masterimage. 
    
\begin{deluxetable}{rcccrrrcccc}
\scriptsize
\tablecaption{Parameters of the SBF Analysis \label{tbl-2}}
\tablewidth{0pt}
\tablehead{
\colhead{Galaxy} &\colhead{$m_1$} & \colhead{exp time} & \colhead{seeing}  &\colhead{$\bar{g}$ } &\colhead{$s(\Delta s)$} &
\colhead{$P_0(\Delta P_0)$} & \colhead{$P_1(\Delta P_1)$} &\colhead{$\bar{m}_{R}(\Delta \bar{m})$} 
& \colhead{A$_R$}  &\colhead{$\bar{m}_R^0(\Delta \bar{m})$} \\
\colhead{}       & \colhead{mag}              & \colhead{sec}   & \colhead{arcsec}      & \colhead{counts}          & \colhead{counts}  & 
\colhead{counts}       & \colhead{counts} & \colhead{mag}   & \colhead{mag}      & \colhead{mag}   \\
\colhead{(1)}    & \colhead{(2)}              & \colhead{(3)}   & \colhead{(4)}      & \colhead{(5)}          & \colhead{(6)} &
\colhead{(7)}    & \colhead{(8)} & \colhead{(9)}      & \colhead{(10)}          & \colhead{(11)} 
}     
\startdata
NGC 59     F1 & 24.30 & 5$\times$450   &  1.4     &   836   &  4166(8) &  34.9(1.7)   &    2.7(0.1) & 27.08(0.06) & 0.06 & 27.02(0.06)\\
           F2 &       &                &          &   706   &          &  33.3(1.8)   &    3.5(0.1) & 27.12(0.06) &      & 27.06(0.06)\\
Scl-dE1    F1 & 24.31 & 5$\times$600   &  1.5     &   101   &  4456(11)&  131.1(16.5) &   18.5(0.4) & 25.96(0.13) & 0.04 & 25.92(0.13)\\
           F2 &       &                &          &    77   &          &  118.8(19.7) &   26.5(0.5) & 26.08(0.17) &      & 26.04(0.17)\\
ESO294-010 F1 & 24.30 & 4$\times$450   &  1.5     &   329   &  3572(7) &  234.3(15.1) &   5.9(0.4)  & 25.00(0.07) & 0.02 & 24.98(0.07)\\
           F2 &     &                  &          &   234   &          &  214.0(13.8) &   10.7(0.5) & 25.11(0.07) &      & 25.09(0.07)\\
ESO540-030 F1 & 24.32 & 5$\times$450   &  1.5     &   193   &  3769(6) &   62.0(4.4)  &   9.7(0.3)  & 26.47(0.07) & 0.06 & 26.41(0.07)\\
           F2 &     &                  &          &    96   &          &   65.8(7.4)  &   24.5(1.0) & 26.41(0.11) &      & 26.35(0.11)\\
ESO540-032 F1 & 24.27 & 5$\times$450   &  1.3     &   123   &  4146(17)&  112.9(17.0) &  12.8(0.5)  & 25.77(0.15) & 0.06 & 25.71(0.15)\\
           F2 &       &                &          &    95   &          &  151.0(28.2) &  16.3(0.6)  & 25.46(0.19) &      & 25.40(0.19)\\
\enddata
\end{deluxetable}

The error of a $P_0$ measurement is dominated by the fitting error and the 
error in the sky value. While the first accounts for 5--9\% in all cases, the latter 
effect depends on $\bar{g}$ going from a negligible 1\% for the fields of NGC 59 
to 18\% for ESO540-032 where the galaxy counts are few. Other possible sources 
of errors are the PSF normalisation and the shape variation of the stellar PSF over 
the CCD area. Using different bright stars as template for the PSF power spectrum 
we found the derived $P_0$ of a field changed by only 1--3\%. Column 7 of 
Table~\ref{tbl-2} gives the values of $P_0$ together with the combined error from 
all discussed sources. Assuming further an accuracy in photometry to $\Delta m_1=0.02$\,mag 
(column 2) and a random extinction error at the South Galactic Pole of 
$\Delta A_R=0.01$\,mag, we get an overall uncertainty for $\bar{m}_R^0$ of 0.06 to 
0.19\,mag (column 11). The error-weighted mean of $\bar{m}_R^0$ for each galaxy 
derived from the two independent fluctuation magitude measurements are given in 
column 2 of Table~\ref{tbl-3}.

\begin{deluxetable}{lrrrrr}
\tablecaption{Mean observed $\bar{m}_R$ values with resulting distances
and absolute magnitudes \label{tbl-3}}
\tablewidth{0pt}
\tablehead{
\colhead{Galaxy} & \colhead{$\bar{m}_R^0$} & \colhead{$(m-M)^0$} & \colhead{D}       &
\colhead{$M_B^0$}  & \colhead{$M_R^0$} \\
\colhead{}       & \colhead{mag}           & \colhead{mag}       & \colhead{Mpc}     & 
\colhead{mag}    & \colhead{mag}  \\
\colhead{(1)}    &\colhead{(2)}            &\colhead{(3)}        &\colhead{(4)}      &
\colhead{(5)}    &\colhead{(6)}
}
\startdata
NGC 59           & 27.04(0.04)             &28.21(0.07)          &4.39(0.15)         &
$-15.30$           &$-16.34$\nl
Scl-dE1          & 25.96(0.10)             &27.13(0.12)          &2.67(0.16)         &
$-9.50$            &$-10.25$\nl
ESO294-010       & 25.04(0.05)             &26.17(0.08)          &1.71(0.07)         &
$-10.67$           &$-11.83$\nl
ESO540-030       & 26.39(0.06)             &27.52(0.08)          &3.19(0.13)         &
$-11.22$           &$-12.02$\nl
ESO540-032       & 25.59(0.12)             &26.72(0.13)          &2.21(0.14)         &
$-10.39$           &$-11.42$\nl
\enddata
\end{deluxetable}

\section{$\bar{M}$ CALIBRATION AND SBF DISTANCES}

We chose to use $R$-band observations (instead of the usual $I$ or $K$) 
to avoid fringing effects on thinned CCDs and to take advantage of the 
relatively darker sky. However, the drawback with this filter at present is the 
missing empirical calibration of $\bar{M}$ in this photometric band. 
Most of the SBF applications have focussed on the $I$-band (e.g.~Tonry 
et al.~1989, 1990; Tonry 1991; Tonry et al.~1997) and $K$-band (e.g.~Luppino 
\& Tonry 1993; Pahre \& Mould 1994; Jensen et al.~1996).  But even working in 
the $I$-band would not help in our case because all galaxies analysed to date are 
high-surface brightness, giant ellipticals, while our objects are low-surface
brightness, dwarf galaxies. The two classes are quite different in almost all of 
their properties, including stellar composition, although these differences are 
by no means well understood (e.g., Ferguson \& Binggeli 1994). In particular, 
$\bar{M}$ strongly depends on the underlying stellar population of a galaxy, and
it would be unclear how to adapt empirical results from E galaxies to dEs. 

To bypass this problem, we constructed a mean $\bar{M}_R$ for dEs
by applying stellar population synthesis models to the local, resolved
dwarf spheroidals. A recent comparison between theory and empirical results 
(Tonry et al.~1997) found very good agreement in the $I$-band when using the model predictions 
of Worthey (1993a, 1993b, 1994). We therefore used Worthey's on-line 
program\footnote{http://astro.sau.edu/$\sim$\,worthey/} to calculate $\bar{M}$ 
values for various photometric bands and different stellar mixtures
composed of a series of single-burst stellar populations. For this procedure 
each population requires a set of input parameters: mean metal abundance, 
age, relative mass weight, and the slope of the IMF. 

To use the program, we first had to estimate the metal abundance 
of the dEs. For this we used the observed [Fe/H]--M$_{V}$ relation 
of Local Group dEs (Da Costa 1994, 1998). 
Absolute visual magnitudes for our dwarfs were estimated 
by combining the extinction-corrected observed $B$ magnitudes, 
the mean $(B-V)$ colour of 0.75 (Bothun et al.~1989), and 
an average distance modulus (m--M)=27.0 for the Sculptor group 
(C97). The resulting magnitudes, in the range $-14.8<M_V<-10.1$, were 
converted into abundances according to the relation 
[Fe/H]$=-0.15\cdot$M$_V-3.45$ (derived from Fig.7 of Da Costa 1998). 
As expected, our dEs are all metal poor systems, with metallicities in 
the range $-1.9<$[Fe/H]$<-1.2$. 

As emphasised by Da Costa (1997), dEs are not single-burst populations like 
a globular cluster. Rather, a diverse and complex
set of star formation histories (SFH) is observed among the local dwarf
spheroidals. Their stellar populations range from old (Ursa Minor) 
and mainly old (e.g.~Tucana, Leo\,II) through intermediate-age episodic 
(e.g.~Carina, Leo\,I) to intermediate-age continuous (e.g.~Fornax). On the 
other hand, Phoenix and LSG3 are classified as dE/Im, because they shows 
similarities to both dwarf spheroidals and dwarf irregulars. These systems 
are dominated by an old metal-poor population
with no evidence for {\em major} star formation activities after the 
initial episode 8--10 Gyr ago. However, both systems have a minor population
of young stars, with ages of about 150 Myr, which makes these galaxies
resemble dwarf irregulars. 

This situation would make us suspect that $\bar{M}_{R}$ for these galaxies is
also diverse. Fortunately, this is not the case. We will show that $\bar{M}_{R}$ 
does not depend sensitively on the star formation history within the observed range. 
This issue will be discussed further in JFB98b. We used Worthey's program to calculate 
$\bar{M}_R$ values for a metal poor old (age $>8.5$\,Gyr) population with 
three different types of SFHs (single burst, episodic, continuous). Seven 
SFHs were modelled with a series of star formation bursts, as given in 
Table~\ref{tbl-4}. The bursts were taken to occur at equal intervals of time between 
the first and the last burst (columns 2--4 of Table~\ref{tbl-4}). The adopted
initial metal abundance was $-1.9$ or $-2.0$. The enrichment per burst
was taken to be constant (column 6), where the value of this constant was 
chosen such that the present-day abundances of the model spanned the range of 
abundances derived above for our galaxies. A Salpeter IMF was adopted 
throughout. The relative mass weight per burst was taken to be constant.
The resulting $\bar{M}_R$ values are listed in column 9.

\begin{deluxetable}{lrcclccrc}
\scriptsize
\tablecaption{Synthetic star formation histories of pure old populations and $\bar{M}_R$ values. \label{tbl-4}}
\tablewidth{0pt}
\tablehead{
\colhead{SFH} &\colhead{Start} & \colhead{Step} & \colhead{End}  &\colhead{Initial [Fe/H]} 
& \colhead{$\Delta$[Fe/H]} & \colhead{Final [Fe/H]} &\colhead{Weight} & \colhead{$\bar{M}_R$} \\
\colhead{}    & \colhead{(Gyr)}              & \colhead{(Gyr)}   & \colhead{(Gyr)}      & \colhead{}          & \colhead{} &
\colhead{}    & \colhead{(\%)} & \colhead{(mag)} \\
\colhead{(1)}    & \colhead{(2)}              & \colhead{(3)}   & \colhead{(4)}      & \colhead{(5)}          & \colhead{(6)} &
\colhead{(7)}    & \colhead{(8)} & \colhead{(9)}
}
\startdata
Burst13            &13  & 0  &  13  & $-1.9$  & 0     & $-1.9$  &  100    &  $-1.136$  \nl
Burst11            &11  & 0  &  11  & $-1.9$  & 0     & $-1.9$  &  100    &  $-1.165$  \nl
Burst9             & 9  & 0  &   9  & $-1.9$  & 0     & $-1.9$  &  100    &  $-1.207$  \nl
Episodical-1.9     &13  & 1.5&  8.5 & $-2.0$  & 0.033 & $-1.9$  &   25    &  $-1.179$  \nl
Episodical-1.2     &13  & 1.5&  8.5 & $-2.0$  & 0.267 & $-1.2$  &   25    &  $-1.138$  \nl 
Continuous-1.9     &13  & 0.5&  8.5 & $-2.0$  & 0.011 & $-1.9$  &   10    &  $-1.177$  \nl  
Continuous-1.2     &13  & 0.5&  8.5 & $-2.0$  & 0.089 & $-1.2$  &   10    &  $-1.142$  \nl 
\enddata
\end{deluxetable}

The modelling reveals $\bar{M}_R$ to be essentially insensitive to the 
changes in SFH for these populations older than $8$\,Gyr. All $\bar{M}$ magnitudes 
are found in the narrow interval $-1.207<\bar{M}_R<-1.136$ with a median value 
$\bar{M}_R=-1.165$. This value can be expected to be a good calibration constant 
for metal poor old populations as observed in Ursa Minor, Tucana, or NGC 205. If we assume that 
the morphological similarities between these LG dwarfs and Scl-dE1 and NGC 59
indicate similar star formation histories, then the median value from the models can 
be adopted for our two sample galaxies. 

The remaining three Scl dwarfs show some indication of a small population of 
young stars as found in Phoenix. To estimate $\bar{M}_R$ for these dwarfs, 
we analysed the old populations (column 1, Table~\ref{tbl-4}) but 
polluted with a 5 and 10 percent (in mass) metal-rich intermediate and 
young population. This second component was introduced in two different 
ways: (1) a constant star formation rate between 1 and 8 Gyr simulated 
with a series of bursts every 1 Gyr, (2) an episodic component with two
bursts at 3 and 6 Gyr. Each burst got equal mass weight and a 
metallicity of [Fe/H]$=-0.225$ (this is the lowest metallicity allowed in 
Worthey's program for bursts younger than 8 Gyr). As in the pure ``old'' cases, 
$\bar{M}_R$ is insensitive to the underlying star formation history of the 
pollution, i.e. continuous or episodical. The important factor which changes $\bar{M}_R$ 
(see Fig.4) is the mass ratio $\mathcal{M}_{\rm young}/\mathcal{M}_{\rm old}$ 
of the two subpopulations. Applying equal weight to the results of all seven 
considered SFHs yields an average empirical relation $\bar{M}_R=-1.165+0.70\cdot
(\mathcal{M}_{\rm young}/\mathcal{M}_{\rm old})$. For ESO294-010, ESO540-030, and ESO540-032, 
we estimate the contribution of the young stars to the total integrated light to 
be less than 3\%. The typical mass-to-light ratio for the generated synthetic 
stellar systems is about 1.5 in the $R$-band, so we consider an old population with 
a young star pollution at the 5 percent level as a fair approximation for 
these three dwarfs. To determine their distances, we will use the mean 
value from our models $\bar{M}_R=-1.13$. 

\begin{figure*}
\centering\leavevmode
\epsfxsize=10cm
\epsfbox{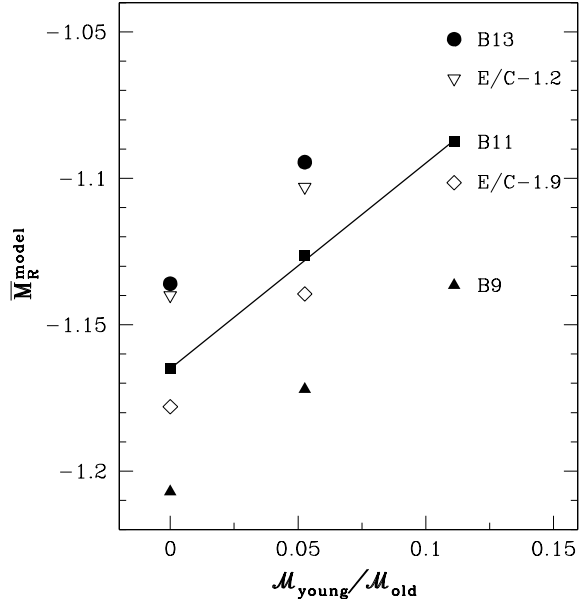}
\caption{
The model $\bar{M}_R$ values for a mainly old ($>$8.5\,Gyr) stellar population
show a systematic trend to fainter magnitudes with an increasing portion of 
young stars. Thereby the slope is independent whether the star formation 
history of the old population is a single burst event (filled circle), continuous 
or episodical (open symbols) as listed in column 9, Table~\ref{tbl-4}. The solid 
line indicates the average linear relation between the mass ratio and $\bar{M}_R$
for our models. 
\label{fig4}}
\end{figure*}

We are very grateful for the use of Worthey's public program which enabled the 
work on $\bar{M}_R$ described above. We note however that star formation histories 
of dEs like Fornax or Carina could not be explored with this program. Both dwarfs 
produced a significant fraction of their metal poor ([Fe/H]$<-1.0$) intermediate 
population between 4 and 8 Gyr (Da Costa 1997). This part of the parameter space, 
i.e.~young ($<8$\,Gyr) and metal poor ([Fe/H]$<-0.225$), is not accepted 
by the model program at present.

As an independent, zero-order check on the intrinsic scatter of $\bar{M}_R$ 
for an old metal poor stellar population, we explored the $\bar{M}$--[Fe/H] 
relation for globular clusters (GC) in $V$ and $I$ (Ajhar \& Tonry 1994) 
for the relevant metallicity range. Our primary interest here is not with 
the absolute values but the observed intrinsic dispersions. Using all 11 GCs 
in their sample with $-1.9<$[Fe/H]$<-1.2$, we derived a mean scatter of 
$\sigma_{V}=0.05$ and $\sigma_{I}=0.07$, respectively. The exclusion of the 
anomalous cluster $\omega$\,Cen reduces the scatter in the $I$ band to $0.06$. 
Assuming that the results for the $R$ band are comparable, we conclude that 
(1) theory {\em and} observations predict a very small variation of $\bar{M}$ 
in this metallicity range, (2) the empirical scatter depends only 
weakly on the photometric passband. 

For the following discussion, we will adopt $\bar{M}_R=-1.17$ (model) for 
our ``dE'' classified dwarfs and $\bar{M}_R=-1.13$ (model) for the intermediate 
type ``dE/Im''. In both cases an error of $\Delta\bar{M}_R=0.06$ (from GCs) is 
adopted.  We regard this as a preliminary answer in the search for the empirical 
calibration of the SBF distance indicator for dEs. Interestingly, the results
from the models are right between the average values of $\bar{M}_I$ and $\bar{M}_V$ 
for metal poor GCs with $-2.03$ and $-0.33$, respectively (Ajhar \& Tonry 1994).

The distances to our galaxies can now be determined in the usual manner:
\begin{equation}
\log(D[Mpc])=0.2\cdot(\bar{m}_R^0-\bar{M}_R-A_R-25).
\end{equation} 
The true (extinction-corrected) distance moduli, the distances in Mpc, and
the resulting absolute total magnitudes in $R$ and $B$ for the five Scl 
dwarfs are listed in Table~\ref{tbl-3}. Also given there are the uncertainties 
in these quantities. We recall here the detailed origin of these errors: 
$(\Delta P_0/P_0)=0.06-0.18$ (fitting, sky, PSF variation), $\Delta m_1=0.02$\,mag 
(calibration), $\Delta \bar{M}_R=0.06$\,mag (modelling), and $\Delta A_R=0.01$\,mag. 
The total SBF distance error for a single measurement is 5 to 10\%.

The new SBF distances and their relevance for the 3D-structure 
of the Sculptor group are discussed in Sec.7 below. First we 
report on a spectroscopic confirmation of the distance for one dwarf.

\section{SPECTROSCOPY}

For an independent proof of group membership for at 
least one of our dEs, we selected our second brightest galaxy 
ESO294-010 for a spectroscopic follow-up. This galaxy was recently 
part of a study on the star formation histories of Sculptor group 
dwarf galaxies (Miller 1996). Miller found that
ESO294-010 is a typical early-type dwarf with no evidence for 
ongoing massive star formation. No \ion{H}{2} regions were discovered 
and the total H$_\alpha$ luminosity is very low. Nevertheless, 
with a sufficiently long integration time we hoped to detect Balmer 
absorption lines in the spectrum which then could be used to measure 
a redshift. This galaxy is even more interesting because 
C97 reported a $2.5\sigma$ detection in \ion{H}{1} at a
redshift of 4450 \kms\ but we measured the galaxy to be only 
2.4 times more distant than M31. 
     
Long-slit spectroscopy was carried out for ESO294-010 
in September 1997. We used the double-beam spectrograph 
(DBS) at the Nasmyth A focus of the 2.3m ANU telescope to observe a 
blue and a red spectrum simultaneously. The detectors were two SITe 
1752$\times$532 thinned CCDs with a pixel size of 15$\mu$m, and a scale 
of 0.9$''$ pixel$^{-1}$ across the dispersion. The wavelength scales were
1.1\AA\, pixel$^{-1}$ in the blue and 0.55\AA\, pixel$^{-1}$ in the red. Beam 
splitter and grating angles were chosen to cover the wavelength ranges 
3500--5500\AA\, and 6000--7000\AA, respectively. The slit was positioned at 
the galaxy centre and aligned along the major axis. The slit length was $6.7$ 
arcmin with a width of 2$''$. A series of four 2000 sec exposures were taken. 
After each science exposure a Ne-Ar lamp was observed for the wavelength 
calibration. 

\begin{figure*}
\centering\leavevmode
\epsfxsize=10cm
\epsfbox{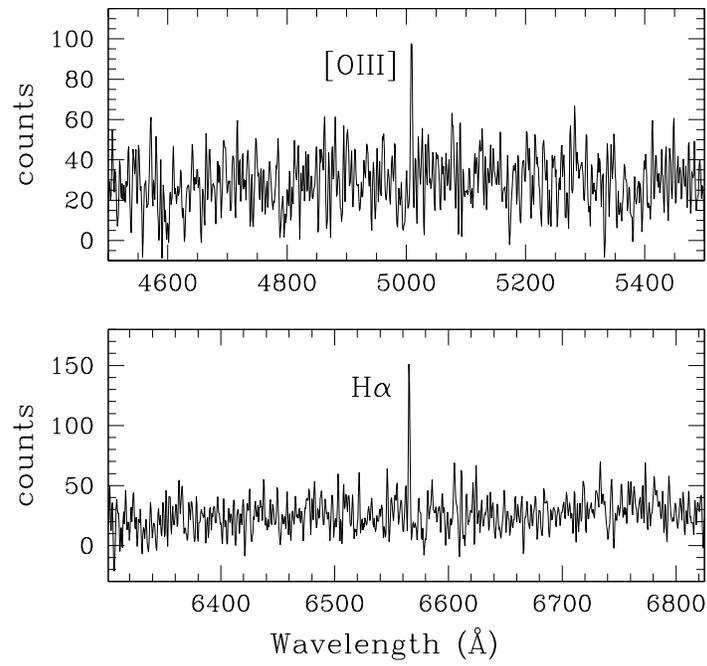}
\caption{Smoothed versions of the blue (top) and red (bottom) spectrum of 
ESO294-010 with a total integration time of 8,000 sec. The centrally located 
[O\,III] and H$_\alpha$ emission lines are the only significant features in 
the spectra. \label{fig5}}
\end{figure*}

IRAF procedures were used for the spectral reduction. The galaxy spectra were 
individually wavelength-calibrated and sky subtracted, and
each set of blue and red spectra were combined to improve the 
signal-to-noise ratio. Fig.~\ref{fig5} shows slightly smoothed versions of the 
important parts of the spectra. No significant Balmer absorption lines are visible. 
However, narrow emission lines at [\ion{O}{3}] $\lambda$5007\AA\,\,and H$_\alpha$ 
$\lambda$6563\AA\,\,are prominent with 5 and 10$\sigma$ significance relative to 
the adjacent continuum rms variations (the two apparent [\ion{S}{2}] lines 
$\lambda$6716\AA\,\, and $\lambda$6731\AA\,\, are not significant). In view of 
Miller's (1996) results mentioned above, this finding was unexpected, even though 
modest amounts of dust and gas in dEs are not uncommon, e.g.~in NGC 185 and NGC 205 
(Hodge 1973, Young \& Lo 1997 and references therein), or NGC 59 (C97). In fact, 
there is a hint of lumpiness in the central part of ESO294-010, clearly seen 
in Fig.~\ref{fig2}, which made us classify this galaxy {\em a posteriori}\/ as an 
intermediate-type dS0/Im (see Sec.2). Tracking down 
the source of the emission, we noticed a circular patch 7 arcsec or 
0.06 kpc across, located 18 arcsec south of the galaxy centre (Fig.~\ref{fig2}). 
This patch coincides in position and extent with the position and width of the 
two emission lines along the spatial axis of the spectra, and is probably
a small \ion{H}{2} region. 
 
The agreement of the velocities derived from the individual lines is excellent, 
with $v_\odot=117 (\pm0.8) $\kms. We suggest that 5\kms\ is a more realistic estimate 
of the velocity error. This low velocity is in accord with our low SBF distance 
for that galaxy, and it will be shown below that it fits nicely into the velocity 
pattern of other Scl group members. 
 
\section{3D-STRUCTURE OF THE SCULPTOR GROUP}

The most important result of our SBF analysis is that all five objects fall 
in the distance range of previously known Scl group members, thus confirming 
their suspected membership as group dwarfs. Moreover, the new dE members help 
to clarify our view of the complex 3D-structure of the group. As it will turn 
out in the following, the notion of a Scl ``cloud'' of galaxies would be a 
better description of this supergalactic structure.

The main members of the Scl group are the six late-type (Sc, Sd, Sm) spirals 
NGC 45, 55, 247, 253, 300, and 7793. Individual distances to these galaxies, 
determined from various methods, are listed in Table~\ref{tbl-5}, column 2. The rms 
uncertainties of the distances are given in parentheses. Most data are taken 
from Puche \& Carignan (1988, Table II). Where more recent measurements are 
available, we have calculated a mean value from the references given in column 
3. One Scl dwarf, the irregular SDIG, which already had a known distance, is 
also included here. To this list we can now add our 5 Scl early-type dwarfs 
with their newly determined SBF distances. 

\begin{deluxetable}{lcllrrl}
\tablecaption{Distances and velocities of galaxies in the Sculptor group region \label{tbl-5}}
\tablewidth{0pt}
\tablehead{
\colhead{Name} &\colhead{D($\Delta$D)} & \colhead{Ref.} & \colhead{v$_\odot$($\Delta$v)}  & \colhead{v$_{GSR}$} & \colhead{v$_{LG}$} & \colhead{Ref.}\\
\colhead{}     & \colhead{Mpc}     & \colhead{}  & \colhead{\kms}  & \colhead{\kms} & \colhead{\kms} & \colhead{}       \\
\colhead{(1)}  & \colhead{(2)}     & \colhead{(3)}   & \colhead{(4)}  & \colhead{(5)} & 
\colhead{(6)} & \colhead{(7)}
}     
\startdata
NGC 55      &  1.66(0.20)    & PC88           & 129(3) & 98  & 111   & DC91       \nl
ESO294-010  &  1.71(0.07)    & this study     & 117(5) & 71  &  80   & this study \nl
NGC 300     &  2.10(0.10)    & Fr92           & 144(1) & 101 & 114   & ESO-LV     \nl
ESO540-032  &  2.21(0.14)    & this study     &        &     &       &            \nl
NGC 247     &  2.48(0.15)    & T87, PC88, F98 & 156(4) & 173 & 211   & DC91       \nl
Scl-dE1     &  2.67(0.16)    & this study     &        &     &       &            \nl
SDIG        &  2.63(0.80)    & L77, H97       & 229(10)& 217 & 238   & C97        \nl
NGC 253     &  2.77(0.13)    & PC88, F98      & 245(5) & 246 & 278   & DC91       \nl
ESO540-030  &  3.19(0.13)    & this study     &        &     &       &            \nl
NGC 7793    &  3.27(0.08)    & T87, PC88      & 227(2) & 226 & 250   & ESO-LV     \nl
NGC 45      &  4.35(1.40)    & PC88           & 463(3) & 488 & 525   & DC91       \nl
NGC 59      &  4.39(0.15)    & this study     & 362(10)& 392 & 432   & C97        \nl
\hline
\enddata
\tablerefs{C97: C\^ot\'e et al.~1997; DC91: Da Costa et al. 1991; 
ESO-LV: Lauberts \& Valentijn 1989; F98: Federspiel 1998 (NGC 247 (m--M)$_0$=27.17, 
NGC 253: (m--M)$_0$=27.36); Fr92: Freedman et al. 1992; H97: Heisler et al.~1997; 
L77: Laustsen et al.~1977; PC88: Puche \& Carignan 1988, and references therein; 
T87: Tammann 1987.}
\end{deluxetable}

Velocities (where available) are also given in Table~\ref{tbl-5}. Heliocentric 
velocities (column 4) are taken from C97, Da Costa et al.~(1991), 
and Lauberts \& Valentijn (1989) for all but one galaxy (ESO294-010) for
which the velocity was established in the present study (see Sec.6).
These velocities were transformed into galactocentric velocities 
(column 5) and velocities relative to the barycentre of the LG (column 6), 
applying the apex vectors given by de Vaucouleurs et al.~(1991) and 
Karachentsev \& Macharov (1996), respectively.

\begin{figure*}
\centering\leavevmode
\epsfxsize=13cm
\epsfbox{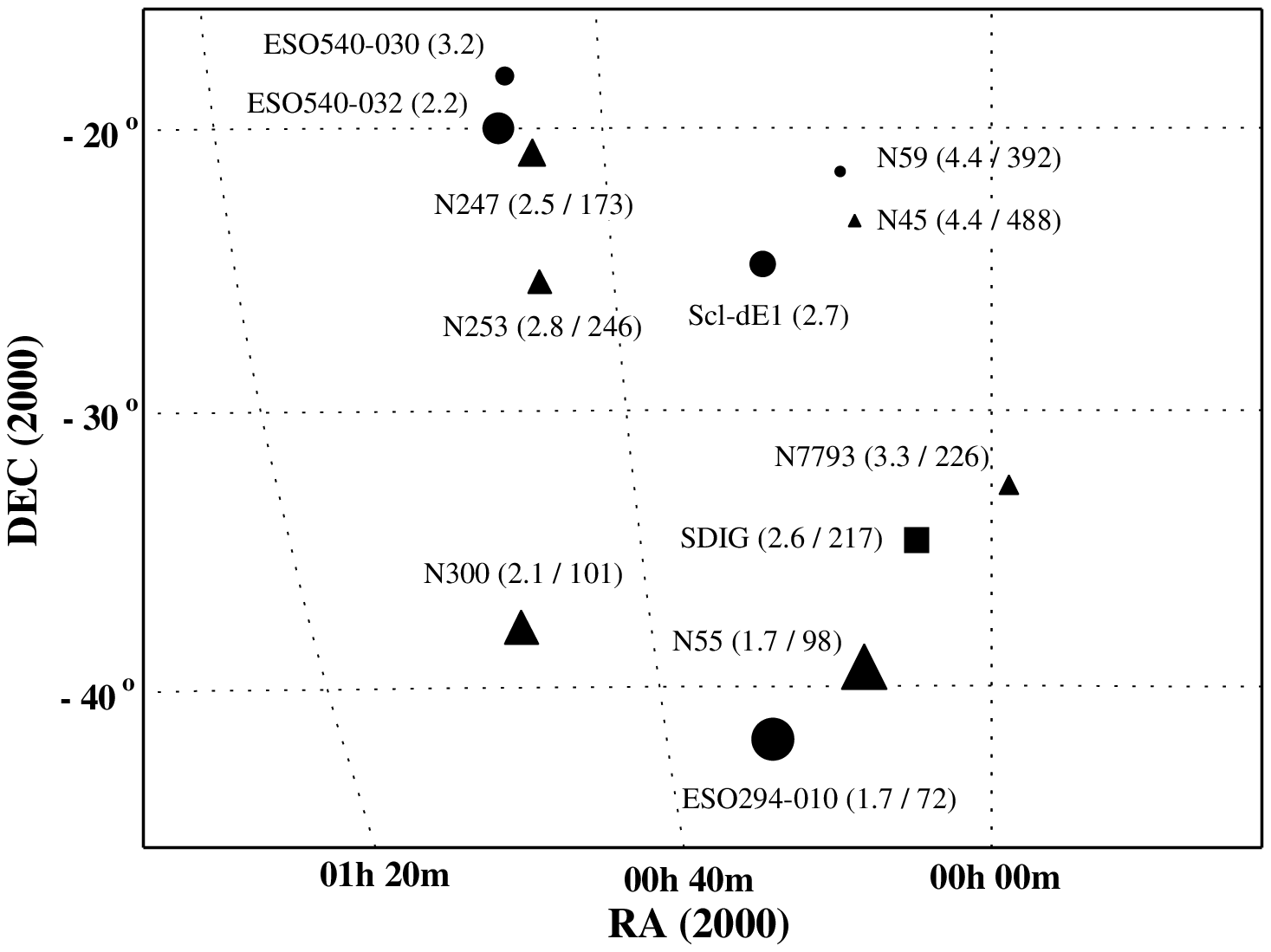}
\caption{Sky distribution of all major Sculptor group members (triangles)
plus SDIG (box) and the dEs (circles) with known distances. Names, 
distances (in Mpc), and galactocentric velocities (if available) 
are indicated. The symbol size is inversely proportional to the 
galaxy distance to simulate a depth effect. \label{fig6}}
\end{figure*}

Fig.~\ref{fig6} is a pseudo-3D plot of the distribution of the 12 Scl galaxies with 
known distances listed in Table~\ref{tbl-5}: it shows the sky distribution of these
galaxies, with the distance of every galaxy indicated by the symbol size
(the distances and velocities are also given as numbers). We first note 
the large depth spanned by the galaxies: distances range from 1.7 Mpc (NGC 55) to 
4.4 Mpc (NGC 45). In fact, Puche \& Carignan (1988) excluded the distant NGC 45
from their study because they insisted on a gravitationally bound Scl group
with a suitably low mass-to-light ratio. Here we relax any dynamical
requirements because, judged from the distribution of galaxies, such a cut
seems artificial. This will become clearer below.

We further note the close match in distance of ESO294-010 with NGC 55 at 
the near end (1.7 Mpc), and of NGC 59 with NGC 45 at the far end (4.4 Mpc) 
of the Scl complex (despite the large distance error for NGC 45). 
It seems very likely that the two dwarfs are bound companions to the major members 
NGC 55 and 45. Both pairs are separated by less than 200 kpc in projection 
and have a radial velocity difference of less than 100 \kms. 
ESO540-030, ESO540-032, and Scl-dE1, on the other hand, seem to be associated 
with NGC 247 and 253 at an intermediate distance of $\approx$ 2.5 Mpc. 
ESO540-032 could be bound to NGC 247. The other two dwarfs are too far 
off for a single companionship to either giant, but a bound substructure 
centred on NGC 247 and 253 is strongly suggested.

The close spatial coincidence of our five dwarf objects with the 
substructure of the Scl group traced out by its main members demonstrates
two things: (1) The SBF method to determine distances to early-type dwarfs 
is reliable and accurate to the claimed level. (2) The trend of ``field'' dEs 
to be satellites of giant galaxies is confirmed, in accord with the 
morphology-density relation for dwarf galaxies (Binggeli et al.~1990). The only 
isolated pure (old-population) dE known so far is the Tucana system in 
the LG (Da Costa 1994, see also Fig~\ref{fig9}. below). The only galaxy classified 
here as a pure dE, Scl-dE1, may also be fairly isolated among our dwarfs.

So far we have not considered the 14 (!) additional dwarf irregular 
Scl group members discovered by C97, because of a lack of distance information. 
However, we can include them in our analysis by using their observed radial 
velocities as distance indicators via the tight velocity-distance relation 
which will be discussed below. Dwarf irregulars are known (and below also 
shown) to avoid high-density regions (cf.~again Binggeli et al.~1990) and 
therefore to possess fairly unperturbed velocities. 

\begin{figure*}
\centering\leavevmode
\epsfxsize=14cm
\epsfbox{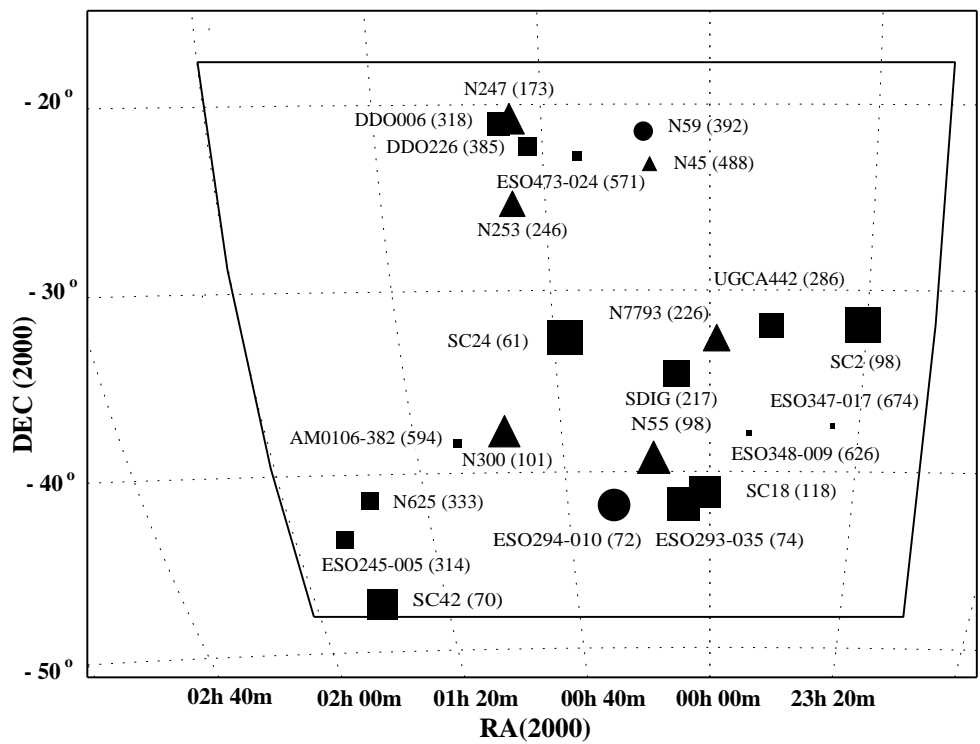}
\caption{Sky distribution of all major group members (triangles), 
dEs (circles), and dwarf irregulars (boxes) with known velocities. Names 
and galactocentric
velocities are indicated. Symbol size is inversely proportional to the 
velocity. Also shown is the survey boundary of the dwarf galaxy searches of
C\^ot\'e et al.~(1997) and JBF98a. \label{fig7}}
\end{figure*}

Fig.~\ref{fig7} is a pseudo-3D plot of the Scl group, like Fig.~\ref{fig6} but now 
with the galactocentric velocity as third dimension. All C97
irregulars are included here, raising the number of Scl galaxies to 23.
We recognize again the NGC 55 subclump on the near side with ESO294-010, and
possibly also SC18 as bound companions. NGC 300 and SDIG are slightly more
distant. Unfortunately, the three dEs associated with NGC 247 and 253
(Fig.~\ref{fig5}) have no velocity and hence are lost here in Fig.~\ref{fig7}. 
DDO006 and 226 with 300\,\kms$<v_{GSR}<$400\,\kms\ must lie behind NGC 247 (173\,\kms).
Likewise, ESO473-024 (571\,\kms) is probably behind NGC 45/59.
On the other hand, a possible pair of irregulars appears in the SE (lower left)
corner with NGC 625 (333\,\kms) and ESO245-005 (314\,\kms). There are some
very nearby dwarf irregulars, such as SC24 (61\,\kms) and SC2 (98\,\kms),
but also some very distant ones, such as ESO348-009 (626\,\kms) and
ESO347-017 (674\,\kms). Overall, the impression is that the dwarf irregulars
are not strongly associated with the main Scl group members but are scattered
over the entire volume in the form of a ``cloud''. This is of course
again in accord with the morphology-density relation for dwarf galaxies
(Binggeli et al.~1990). The whole Scl group seems to have the shape
of a cigar of length 3 Mpc and thickness 1 Mpc, which we see pole-on because
the LG lies near its end. A much better view of this can be gained 
if the cigar is turned to the side by plotting the distances linearly.
For this we have first to transform velocities to distances via a Hubble diagram.

\begin{figure*}
\centering\leavevmode
\epsfxsize=12cm
\epsfbox{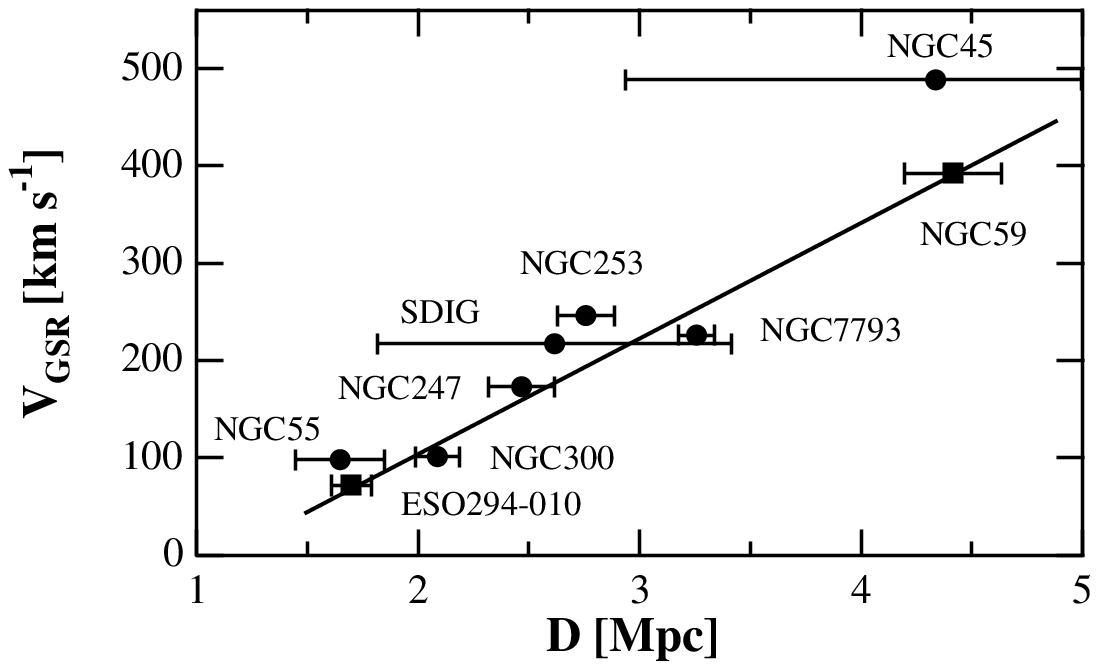}
\caption{Hubble diagram for the 9 Scl group galaxies with known velocities
{\em and}\/ distances (from Table~\ref{tbl-5}). The two filled squares 
are the new entries, ESO294-010 and NGC 59. All points are shown with 
$\pm1\sigma$ distance error bars. Velocity errors are small for all galaxies 
($<5$\%) and not indicated here. The solid line is the best linear ML fit to the 
data. 
\label{fig8}}
\end{figure*}

In Fig.~\ref{fig8} we show the Hubble diagram for the total of nine 
galaxies with known distances and velocities, to which we have contributed 
ESO294-010 and NGC 59. The velocities ($v_{GSR}$) are relative to the centre 
of our 
Galaxy. The major Scl group members show a tight linear relation between 
velocity and distance. Such a trend was already seen by  Puche \& Carignan 
(1988) and is now confirmed and amplified by the two new data point from 
the dEs (filled squares). A maximum likelihood linear fit to the data yields: 
\begin{equation}
v_{GSR}[{\rm km~s}^{-1}]= 119(\pm7) \cdot D [{\rm Mpc}]\,\,-136(\pm14)\,, 
\end{equation}
where $v_{GSR}$ is taken to be the independent variable and the distance 
errors are given in Table~\ref{tbl-5}.

The mean scatter of the residuals about the line is remarkably small, with 
$\sigma_{D}=0.34$\,Mpc (0.23\,Mpc without NGC 45) and $\sigma_{v_{GSR}}=
40$\,\kms (27\,\kms without NGC 45). Large peculiar velocities due to 
internal group dynamics are clearly absent. This strongly suggests that our 
Scl galaxies form a ``cloud'' (e.g. de\,Vaucouleurs 1975, Tully 1982, 
Tully \& Fisher 1987), meaning a supergalactic structure of higher-than-average 
galaxy density which, in contrast to a ``group'', is gravitationally 
{\em unbound}. However, bound substructures, e.g.~defined by one giant 
galaxy plus a swarm of dwarf companions, are often found embedded in such 
a cloud. This also seems to be the case here.

The slope of the Hubble relation of Equ.5 is very steep (119\,\kms\,Mpc$^{-1}$).
A plausible reason for this deviation from the global Hubble law ($H_0 \approx 
60 \pm$ 10 \kms\,Mpc$^{-1}$) is the gravitational influence of the LG. The 
deceleration of the local expansion field (for $D<$ 4 Mpc) by the LG was 
nicely recovered by Sandage (1986, 1987) who used his local galaxy data to 
put constraints on the mass of the LG. However, an attempt to fit one of 
Sandage`s template curves for the local Hubble diagram (1986, his Fig.~2) to 
our data (Fig.~\ref{fig8}) proved unsuccessful. At present we do not understand the 
dynamics underlying the Scl cloud velocity field, and a dynamical analysis is 
beyond the scope of this paper.

\begin{figure*}
\centering\leavevmode
\epsfxsize=17cm
\epsfbox{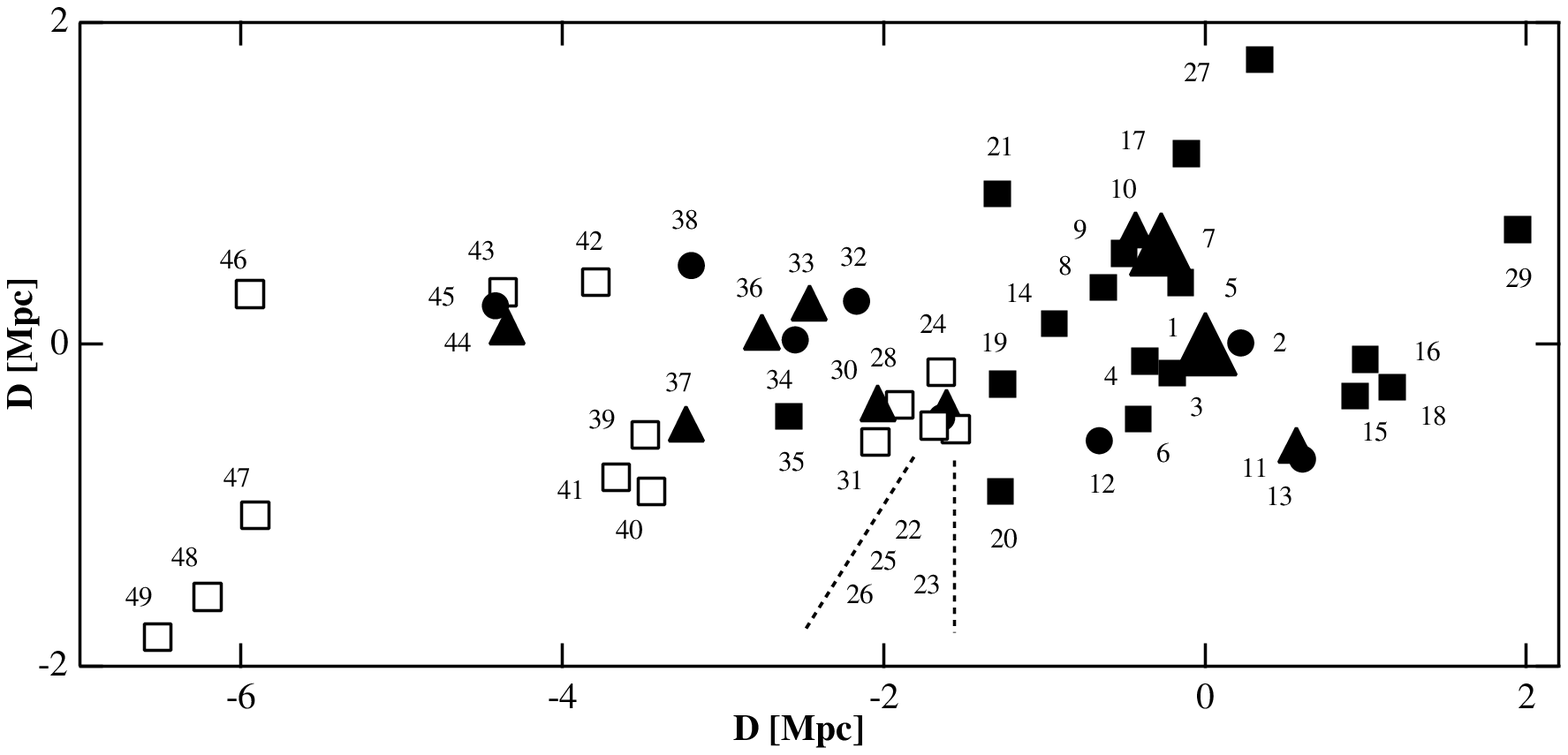}
\caption{A map of the Local Group and its extension towards the Sculptor 
group region out to a distance of 6 Mpc, projected onto the supergalactic 
plane. Filled symbols are galaxies with measured distances. Circles: dwarf 
elliptical galaxies. The five new dE galaxies are \#25, 32, 34, 38 and 45.
Triangles: spiral galaxies. The two larger triangles indicate the positions of 
the Milky Way (at the origin) and M31 with their systems of close dwarf companions.
Boxes: dwarf irregular galaxies. Open boxes are the dwarf irregulars detected by 
C\^ot\'e et al.~(1997) with distances reckoned from their velocities through the 
Hubble diagram (Fig.~8). The dashed lines show a zoom-in of the aggregate at NGC 55. 
The names of the galaxies are, 1: MW \& companions, 2: LeoI \& II, 3: NGC 6822, 
4: Phoenix, 5: DDO210, 6: SagDIG, 7: M31 \& companions, 8: IC1613, 9: LSG3, 
10: M33, 11: NGC 3109, 12: Tucana, 13: Antlia, 14: WLM, 15: Sex\,A, 16: Sex\,B, 
17: IC10, 18: GR8, 19: ESO407-018, 20: IC5152, 21: Pegasus, 22: NGC 55, 23: SC42, 
24: SC24, 25: ESO294-010, 26: ESO293-035, 27: UGC-A86, 28: SC2, 29: Leo\,A, 
30: NGC 300, 31: SC18, 32: ESO540-032, 33: NGC 247, 34: Scl-dE1, 35: SDIG, 36: 
NGC 253, 37: NGC 7793, 38: ESO540-030, 39: UGCA442, 40: ESO245-005, 41: NGC 625, 
42: DDO006, 43: DDO226, 44: NGC 45, 45: NGC 59, 46: ESO473-024, 47: AM0106-382, 
48: ESO348-009, 49: ESO347-017. \label{fig9}}
\end{figure*}

Because the velocity-distance relation of Equ.5 is so tight, we can use it
empirically to obtain approximate distances to those Scl galaxies which have
only velocities, i.e.~the dwarf irregulars of C97. Based on these 
distances we show, in Fig.~\ref{fig9}, the distribution of {\em all}\/ 
Scl cloud members in projection on to the supergalactic plane (SGP). 
This provides us with a side view of the cigar-shaped Scl cloud, as it was 
described above. The scatter of the third dimension here, the height above/below
the SGP, is quite small ($\sigma_{z}=0.53$\,Mpc). We have also included LG 
``members'' and galaxies at the outer fringes of the LG in this plot 
(data taken from van den Bergh 1994), to demonstrate that the Scl cloud 
and the LG may actually form {\em one} structure. This conjecture was 
raised before by Binggeli (1989). There is a marked asymmetry in the 
distribution of galaxies within a distance of 1.5 Mpc. Aside from
the close companions to our Galaxy and M31, most LG dwarf members and 
suspected members (IC1316, WLM, NGC 6822, Phoenix, Tucana, SagDIG, DDO210,
Pegasus, IC5152, ...) form a southern extension of the LG -- A
bridge across to the Scl cloud (see also Fig.2 in Binggeli 1989).

The whole structure can be described as a prolate (6 $\times$ 2 Mpc) cloud
of (unbound) dwarf irregulars, in which a number of denser groups are
embedded, each containing one or more massive galaxies plus a swarm 
of (bound) dwarf companions which tend to be of early type. Certainly our 
Galaxy and M31 with their satellites form such condensations -- each alone, 
and together as the LG. In the Scl region we note again the three 
subclumps mentioned before: centred on NGC 55 (plus NGC 300?), NGC 247/253, 
and NGC 45. NGC 7793 may define a fourth subclump or may be member of the 
first (Davidge 1998).  We note again the position of our five early-type 
dwarfs in this map: they tend to mark the condensations. 

The dynamics of these Scl subsystems with respect to each other, and with
respect to the LG in particular, is not clear at present. We note that 
the major Scl group members are much less massive than M31 or our Galaxy, and 
that M31 and the Galaxy lie at the extreme end of the prolate cloud,
but we do not know if this is physically significant.
Finally, it is worth mentioning that the Scl-LG cloud is 
part of a yet larger structure -- a prolate cloud that stretches
out to the Coma\,I cluster at $D \approx$ 15 Mpc, in the direction of the
Virgo cluster. The feature is indeed called the ``Coma-Sculptor cloud''
(Tully \& Fisher 1987, see plate 14).

\section{SUMMARY AND CONCLUSIONS}

We have shown that the Surface Brightness Fluctuation method can be 
successfully used as a distance indicator for low surface brightness dwarf 
elliptical galaxies as faint as $M_{B}=-9.5$. The application to the dwarf 
ellipticals rather than normal ellipticals has two advantages: (1) there
are many dEs in the LG and its vicinity, and independent distances to
these nearby dEs can be used to calibrate the absolute fluctuation magnitude 
$\bar{M}_R$, (2) Through HST studies, we have detailed information 
about the variety of stellar content and star formation history of this 
galaxy type. However, most of them are dominated by an old metal poor 
stellar population. In this situation, first results from stellar population 
synthesis models suggest that $\bar{M}_R$ is relatively insensitive to the SFH.

The internal fluctuation error for our dEs is 0.04-0.12\,mag which includes 
the results from comparison of two different fields of the same galaxy.
A small systematic uncertainty for $\bar{M}_R$ comes from the unknown 
stellar content of the galaxy. For example, $\bar{M}_R$ gets 0.12\,mag 
fainter when going from a pure old population (all stars older 
than 8 Gyr) to an old population contaminated by 10\% of young population. 
This trend was taken into account here by applying appropriate calibration 
constants to the different morphological galaxy types. The overall 
distance error we derive is between 4 and 7\%. Although a  more detailed 
analysis is required to follow up the population 
issue, a very satisfactory hint at the reliability of the theoretical values 
comes from the fact that all dEs turned out to be close companions of Scl 
group members for which independent distance measurements exist. These 
parent galaxies cover the full 3\,Mpc depth of the Scl group. 

Based on the results for our five dEs, we further reviewed the 
kinematics and the 3D-distribution of the Scl group. Surprisingly, the 
Hubble flow at the group shows no evidence for disturbance due to 
gravitational interaction between the main group members but is possibly 
decelerated by the tidal force of the nearby, massive LG leading to a 
large local Hubble constant of $119$\kms\,Mpc$^{-1}$. Taking advantage of 
the tight empirical velocity-distance relation we determined distances
to dwarf irregulars in the Scl region and compiled distances from the 
literature for galaxies in the local volume to illustrate that the LG 
and Scl group are not isolated in space but are part of a large prolate 
cloud (6 $\times$ 2 $\times$ 1 Mpc$^3$ in size) of dwarf irregulars well 
defined within the supergalactic plane. A number of groups are embedded
in this structure, each containing one or more massive galaxies 
accompanied by a population of bound dwarf satellites which tend to 
be of early type. 

Finally it is interesting to note that four of our dE galaxies 
were already known prior this study. It is their membership of the
Scl group which was uncovered here. This is a clear signal that the 
search for nearby dwarf elliptical galaxies should not only be 
focussed on the discovery of new, very low surface brightness 
galaxies but also on a careful follow-up of already known 
galaxies which appear nearby from their morphology. For such
follow-up work, the SBF method is a powerful tool for measuring  
the distance of a galaxy. For the future, this opens up the possibility 
to measure accurate distances to all known dEs in the distance 
range between 1 and 10\,Mpc without the expensive requirement of resolving 
the stellar system into stars. This will allow us to establish 
background-free dE samples, to explore the galaxy content of medium and 
low-density regions of the nearby universe, and to study the physical 
properties of dEs in greater details. Furthermore, the existence (or 
non-existence) of isolated early-type dwarfs can be established as a
constraint on the formation and evolution of galaxies.  

\acknowledgments 

We are grateful to John Tonry and Alan Dressler for very helpful
advice, to the referee who made some important suggestions, and to 
Sylvie Beaulieu for doing test observations of NGC 5128 which were 
used to establish the SBF method on low-surface brightness regions. 
Two of us (HJ and BB) thank the Swiss National Science Foundation for 
financial support. HJ thanks further the Freiwillige Akademische 
Gesellschaft der Universit\"at Basel and the Janggen-P\"ohn Stiftung 
for their partial financial support. This project made use of the 
NASA/IPAC Extragalactic Database (NED), which is operated by the Jet 
Propulsion Laboratory, Caltech, under contract with the National Aeronautics 
and Space Administration.

\end{document}